\documentclass[fleqn,usenatbib]{mnras}

\usepackage{newtxtext,newtxmath}
\usepackage{multirow}

\usepackage[T1]{fontenc}

\DeclareRobustCommand{\VAN}[3]{#2}
\let\VANthebibliography\thebibliography
\def\thebibliography{\DeclareRobustCommand{\VAN}[3]{##3}\VANthebibliography}

\usepackage{graphicx}	
\usepackage{amsmath}	
\newcommand{\angstrom}{\textup{\AA}}

\title[AI Techniques in Wide Field Astronomy]{AI Techniques for Uncovering Resolved Planetary Nebula Candidates from Wide-field VPHAS+ Survey Data}

\author[Sun et al.]{
Ruiqi Sun,$^{1}$
Yushan Li,$^{2}$\thanks{liys9378@gmail.com}
Quentin Parker,$^{3}$ \thanks{quentinp@hku.hk}
Jiaxin Li,$^{1}$
Xu Li,$^{1}$
Liang Cao,$^{1}$
and Peng Jia$^{1,4}$ \thanks{robinmartin20@gmail.com}
\\
$^{1}$College of Electronic Information and Optical Engineering, Taiyuan University of Technology, Taiyuan, 030024, China\\
$^{2}$The Department of Physics, The University of Hong Kong, Hong Kong SAR, China\\
$^{3}$The Laboratory for Space Research, The University of Hong Kong, Hong Kong, 999077, China\\
$^{4}$Peng Cheng Lab, Shenzhen, 518066, China
}

\date{Accepted XXX. Received YYY; in original form ZZZ}

\pubyear{2015}

\begin{document}
\label{firstpage}
\pagerange{\pageref{firstpage}--\pageref{lastpage}}
\maketitle

\begin{abstract}
AI and deep learning techniques are beginning to play an increasing role in astronomy as a necessary tool to deal with the data 
avalanche. Here we describe an application for finding resolved Planetary Nebulae (PNe) in crowded, wide-field, narrow-band 
H$\alpha$ survey imagery in the Galactic plane. PNe are important to study late stage of stellar evolution 
of low to intermediate-mass stars. However, the confirmed $\sim 3800$ Galactic PNe fall far short of the numbers 
expected. Traditional visual searching for resolved PNe is time-consuming due to the large data size and areal coverage of modern 
astronomical surveys, especially those taken in narrow-band 
filters highlighting emission nebulae. To test and facilitate more objective, reproducible, efficient and reliable trawls for PNe 
candidates we have developed a new, deep learning algorithm. In this paper, we applied the algorithm to several H$\alpha$ digital 
surveys (e.g. IPHAS and VPHAS+). The training and validation dataset was built with true PNe from the HASH database. After 
transfer learning, it was then applied to the VPHAS+ survey. We examined 979 out of 2284 survey fields with each 
survey field covering $1^\circ\times 1^\circ$. With a sample of 454 PNe from the IPHAS as our validation set, 
our algorithm correctly identified 444 of these objects (97.8\%), with only 16 explicable `false' positives. 
Our model returned $\sim$20,000 detections, including 2637 known PNe and many other 
kinds of catalogued non-PNe such as HII regions. A total of 815 new high-quality PNe candidates were found, 31 of which were 
selected as top-quality targets for subsequent optical spectroscopic follow-up. Representative preliminary confirmatory spectroscopy results are presented here to  demonstrate the effectiveness of our techniques with full details to be given in paper-II.
\end{abstract}




\section{Introduction}
\label{sec:intro}

Until the advent of digital imagery, celestial sources were "discovered" by visual examination, first of the sky 
itself and then from photographic plates. Eventually such plates were digitised by microdensitomers like the APM 
\citep{Irwin1994} and SuperCOSMOS \citep{2001MNRAS.326.1279H} measuring machines that allowed for automatic image 
detection algorithms to be developed - so called Image analysers, e.g. \citep{Stobie1986}. These powerful 
techniques allowed for both image detection and separation of stars and Galaxies across vast areas of the sky 
from application to digitised Schmidt telescope plates. These were of entire surveys of the Northern and Southern 
skies in various optical pass-bands such as B, R and I. Such techniques included automatic following of the 
variable sky backgrounds across these survey plates and image deblending. Such sophisticated software worked well 
for regions outside the Galactic plane. Image crowding and extensive areas of Galactic emission from ionised 
gas becomes a serious issue only inside the Galactic plane.

The advent of the first wide-field narrow-band surveys of the Galactic plane in H$\alpha$ such as the 
SHS \citep{2005MNRAS.362..689P} and VPHAS+ \citep{2014MNRAS.440.2036D} in the South and IPHAS \citep{2005MNRAS.362..753D} 
in the North, enabled the contrasted detection with the off-band equivalent broad-band red images of various kinds 
of emission line star and resolved, isolated and extended nebulosities. The character and identification of these 
extended nebulosities in particular, which may comprise HII regions, Wolf-Rayet shells, Supernova remnants, 
Planetary nebulae (PNe), str\"{o}mgren zones, general and extensive ionised ISM and more, are varied and complex. 
This makes automatic detection and classification difficult and beyond the capability of traditional image analysers 
of the connected pixel image data type.  As a result, visual scrutiny of these H-alpha survey imagery from both 
digitised wide-field UKST tech-pan films for the SHS \citep{2005MNRAS.362..689P} and the CCD imagery from IPHAS 
and VPHAS+ was resorted to once more. This was to discover and identify discrete Galactic emission Nebulae, a 
process that is inevitably subjective, inefficient and imperfect. Nevertheless, it has resulted in large numbers 
of new Galactic PNe being discovered in both the Southern Sky \citet{2006MNRAS.373...79P} and \citet{2008MNRAS.384..525M} 
from the SHS and Northern Sky from IPHAS, \citet{2014MNRAS.443.3388S} where amateur scrutiny of the on-line survey 
data has also played an increasing role, e.g. \citet{2022A&A...666A.152L}. Another PNe candidate discovery method 
is to select candidates from the photometric results of compact H$\alpha$-emitting sources, or using a color-color 
diagram of IPHAS (r’ - H$\alpha$) versus (r’ - i’) and 2MASS (J - H) vs. (H - Ks), which is suitable for compact 
PNe smaller than 5 arcsec. For example, \cite{2009A&A...504..291V, 2009A&A...502..113V} found thousands of candidates 
with this method. However, most have been shown to be emission line stars of various kinds so this process for 
finding PNe is extremely inefficient.

In this work, we have investigated the use and power of deep-learning techniques to replace this retrograde visual 
examination. We have developed bespoke algorithms to automatically process these complex Galactic Plane 
narrow-band H$\alpha$ images to see if they can be employed to systematically detect discrete emission 
nebulosities with a particular emphasis on Planetary Nebulae (PNe). 

This paper is organised as follows: In Section \ref{sec:class} we introduce PNe as an object class and their associated 
identification problems. In Section \ref{sec:surveys}, we introduce the IPHAS and VPHAS+ H$\alpha$ survey projects 
as the data used in this work. In Section \ref{sec:method}, we detail the methods used, including algorithm 
construction and data processing methods. In Section \ref{sec:results}, we demonstrate the excellent detection 
performance of our new model using different indicators and discuss our new nebula target catalogue. In Section 
\ref{sec:sample analysis}, we provide observations and analysis for several newly discovered nebulae as examples and 
conduct relevant discussions. Finally. in Section \ref{sec:conclusions and prospects}, we summarize and discuss 
the current work and prospects for the future.

\section{Planetary nebulae as a class}
\label{sec:class}
PNe are the gaseous, envelopes ejected by 1-8 $\rm M_\odot$ stars towards the ends of their lives that are ionized 
by the radiation of their heating central stars (CSPN) that are well on the way to becoming white-dwarfs. They are some 
of the most important and complex astrophysical objects to understand, especially as they probe mass-loss physics and 
stellar nucleosynthesis products  \citep{1995PhR...250....2I}. Their progenitor stars constitute the majority of low- to 
intermediate-mass stars and are one of the primary contributors of ejected, enriched material into the interstellar medium. 
With many strong diagnostic emission lines in their spectra, abundance, densities, temperatures, ionisation and, via 
photo-ionisation modeling, central star properties can all be determined as well as more general properties such as 
Galactic abundance gradients \citep{2010ApJ...724..748H,2003A&A...397..667M}. For a recent, comprehensive PNe review 
see \cite{2022PASP..134b2001K}. All currently confirmed Galactic PNe  are contained in the ``Gold Standard'' 
HASH database\footnote{HASH: online at \url{http://www.hashpn.space}. HASH federates available multi-wavelength imaging, 
spectroscopic and other data for all known Galactic and Magellanic Cloud PNe.}, 
e.g. \citet{2016JPhCS.728c2008P,2022FrASS...9.5287P}.

\subsection{Problems of identification}
After 260 years of observation and decades of theoretical studies, only $\sim 3880$ PNe have been found in the Galaxy 
\citep{2016JPhCS.728c2008P}. This is much lower than the best theoretical estimations of between $\sim 6600$ to $\sim 45000$ 
depending on stellar population synthesis model assumptions, e.g.  \citet{2006ApJ...650..916M}. Better estimates of the 
actual Galactic PNe population via more complete discovery processes can help rule out different PNe formation schemes 
and on whether they form from single stars or have to emerge from binary systems \citep{DeMarco2009}. Despite their short 
lifetime of typically 21,000 $\pm$ 5,000 years \citep{2013A&A...558A..78J} (but see \citep{2022ApJ...935L..35F}), PNe 
experience drastic changes as they evolve, resulting in diverse properties of morphology, size, surface brightness 
distribution, spectral characteristics etc, e.g. see  \citet{2022FrASS...9.5287P} for a recent review of the identification 
issues at play. More complete samples of the underlying PN population are crucial to understanding their role but 
simultaneously, these broad variation in observed properties also increases the difficulty in discovering them in a fair 
and unbiased way. Indeed, many PNe are faint and hidden in the dense star fields of the Galactic bulge and plane, which 
brings significant challenges to their discovery in such environments. 

Traditional PNe detection methods are inefficient once size and surface brightness fall below 
certain limits and the surrounding environment becomes too complication in terms of star density or extensive unrelated 
emission giving rise to severe selection effects. Standard, automated, model-fitting 
methods also have limitations when applied to targets with complicated shapes and low surface brightness. This has 
motivated us to develop a novel automated machine learning (ML) algorithm to quickly and reliably find PNe in wide-field 
CCD imagery. In recent years, ML has demonstrated strong image recognition capabilities and made significant contributions 
in astronomical survey data processing. For PNe, \cite{2020Galax...8...88A} designed a deep 
transfer learning algorithm for PNe identification and morphological classification, using multiple 
sets of optical and infrared data. They tested various deep learning networks, most of which 
achieved an accuracy rate of only $80\%$. We chose H$\alpha$ survey data for searching for this work because 
traditional visual investigation, based on these new narrow-band surveys, has accumulated a large number of reliable 
targets for training. Moreover, the latest H$\alpha$ survey of the Southern Galactic plane, VPHAS+ 
\citep{2014MNRAS.440.2036D}, has so far been under-exploited and has the greatest current potential 
for discovery of PNe across the most dense and crowded parts of the Galactic plane, including the Bulge. VPHAS+, 
undertaken on the VST in Chile, has higher-resolution than the SHS and similar resolution to the equivalent
IPHAS survey in the North but like IPHAS only samples the inner 5~degrees either side of the mid-plane in Galactic 
latitude. The VPHAS+ survey ran for 7 seasons until mid-August 2018 and obtained images of 91.6\% 
of its planned footprint. Large-scale search work for new nebulosities has not really begun. This work focuses on 
searching for PNe in the VPHAS+ data of dense star fields towards the Galactic bulge with new, powerful deep learning 
methods we have developed.

\subsection{Application of deep learning techniques to PNe identification}
A new deep learning technique, known as a transformer model, is gaining popularity for a wide range of applications 
\citep{vaswani2017attention}. This model relies on something called an "attention mechanism," which consists of multiple layers of self-
attention modules and feedforward neural networks. These layers capture information at different scales and levels of abstraction in the 
image, enabling the model to recognize complex patterns, structures, and relationships within the image, which allow the transformer to 
process information in a unique way. Unlike traditional deep convolutional neural networks, which have limitations on the area of data they 
can analyze at once, transformer models are different. The convolutional neural networks process images at a small, fixed-size window on a 
big picture. They can see only what's inside that window (known as receptive field), and this can be limiting when they are used to process 
images which have large size and contain targets with complex structures. The transformer could process the entire image all at once. They 
don't have these fixed windows, so they can understand how different parts of the picture relate to each other, no matter how near or far 
they are. In simple terms, the transformer has a more versatile and powerful tool to process large images.\\

 The transformers exhibit better detection performance and generalisability. However, a challenge 
 of using a transformer model for object detection is how to solve the problem of model complexity. In 
 \cite{2023AJ....165...26J}, this technology was first applied to the detection of strong gravitational lenses at 
 the scale of galaxy clusters. However, in this architecture, convolutional models are still retained to solve the 
 problem of model complexity by extracting the initial features of the targets through the "ResNet" network 
 \citep{he2016deep}. This inevitably leads to the loss of some target information and a decrease in the 
 performance of small target detection. Compared with the large-scale effects of strong gravitational lenses, 
 Galactic nebulae typically exhibit more diverse shapes and a wider range of scales. Swin-Transformer 
 \citep{2021arXiv210314030L}, is an improvement of the transformer that solves the problem of model complexity 
 while retaining its powerful feature extraction capabilities. Therefore, unlike the method employed by 
 \cite{2023AJ....165...26J}, the model we use is a Swin-Transformer model based on the Mask R-CNN architecture 
 \citep{he2017mask}. Specifically, we replace the feature extraction network in Mask R-CNN with a transformer 
 network, replace the ResNet module with a Swin-Transformer, and construct a new 'bespoke' deep learning object 
 detection method for PNe.

The model we constructed is a data-driven algorithm, so selecting the appropriate dataset is very important. To 
begin with we made use of the IPHAS H$\alpha$ photometric survey of the Northern Galactic plane 
(\citet{2005MNRAS.362..753D} and see below). 
Through the efforts of \cite{2014MNRAS.443.3388S} in particular, a large number of high-quality PNe targets have 
already been found and confirmed in the IPHAS survey images. Therefore, we used this PNe catalog and IPHAS survey images to 
construct datasets to use in this work. We used 1137 individual images of PNe from the IPHAS survey and the 
corresponding data catalogue to train our model. We used a further 454 images to validate the model, and 
determined the detection ability of the model by comparing the model's output results with the catalog data of 
these images. We use the validation data set to test the performance of our method. Our results show that our method could detect $97.8\%$ of all known PNe while $96.5\%$ of the detected PNe are true PNe according to HASH. This shows that our algorithm can not only independently find almost all the known PNe 
targets in the survey data but also has a very high accuracy rate. This means it can reduce a lot of manual 
candidate vetting and confirmation. Confidence in such automated process is of great significance when trawling 
through wide-field large-scale survey imagery.

VPHAS+ is another high-resolution H$\alpha$ photometric CCD survey of the southern galactic disk and bulge (see 
below). As it has not been yet systematically searched by human eyes or AI techniques it  provides 
 large potential for new discovery, especially in the dense, rich bulge region. Considering the 
overall similarity between the IPHAS and VPHAS+ datasets, we directly applied the model trained on 
the IPHAS dataset to also search in VPHAS+. We examined 979 fields covering 2284 square degrees of VPHAS+ data. 
This amounts to about 4.5 million processed images with size $512\times 512$ square pixels 
from which we obtained $\sim$20,000 detections. We compared 
these detections with existing catalog and resulted in 2637 known PNe and other kinds of nebulae. We then 
inspected the rest of the images visually, eliminating bright stars, CCD issues and large-scale diffuse nebulae. 
After this we found 815 new high-quality candidates for follow-up.

We selected 31 of the most promising candidates for undertaking confirmatory spectroscopic observations 
on the SAAO 1.9~m telescope in Sutherland, South Africa in June 2023 using the SpuPnic spectrograph. These 
results are reported in detail in Yushan Li et al., paper~II but some preliminary, representative, confirmatory, reduced 1-D spectra are shown in section 5.2.  A more comprehensive search for nebulae 
targets in the VPHAS+ data (referring to further visual inspection in fields with more offset pixels) is also underway.  
We believe that the attention mechanism-based nebula target detection 
model we have constructed has not only achieved excellent results in nebula target detection, discovering many new 
targets, but also has great promise for detection of other astronomical targets in other types of wide field survey  
data  such as with the CSST\footnote{\url{https://nao.cas.cn/csst/}} and LSST\footnote{\url{https://www.lsst.org/}}.

\section{H$\alpha$ Surveys}
\label{sec:surveys}

H$\alpha$ surveys can be used to search for many types of H$\alpha$-emitting sources referred to earlier, including 
both compact and resolved object types. Such H$\alpha$ surveys generally include both the H$\alpha$ and red broad-band 
`r' exposure (which includes H$\alpha$ in the band-pass). The H$\alpha$ line is isolated with a narrow band optical 
interference filter (typically 75-100\AA wide) centered around 6563 $\angstrom$ and includes the H$\alpha$ emission 
line and the adjacent [N~II] emission lines. These lines, in various ratios, are characteristic of PNe and many other 
types of emission object. The r band includes the H$\alpha$ line in the bandpass and because the r filter is broad 
it can be effectively used as an off-band comparison. During observation, the strategy is to adjust 
the exposure time of both bands, so that the limiting magnitude in both is approximately equal. 
By comparing the two bands, PNe with strong emission lines (including [NII]) near H$\alpha$ can be 
detected by the algorithm. Subsequently, other targets such as emission line stars 
and large-scale diffuse nebulae can be excluded with visual inspection, based on whether the target is 
discernible, its shape, color, and other features.

In this work, we used the IPHAS and VPHAS+ surveys as our primary datasets. These are two important CCD based H$\alpha$ 
surveys covering the Northern and Southern Galactic planes respectively. Their basic properties are listed in 
Table \ref{tab:survey}. Figure \ref{fig:surveys} shows the same PNe Abell 47, PHR J1843-0232, PHR J1843+0002 
(HASH ID\# 352, 2469, 2478) taken from an area of overlap between these two CCD surveys and also the same PN image 
from the earlier SHS photographic H$\alpha$ to demonstrate the difference of resolution and quality of 
these three surveys. All surveys have similar sensitivity for H$\alpha$ emission but with morphological detail 
more evident in IPHAS and VPHAS+.

\subsection{SHS}
\begin{figure}
\includegraphics[width=\columnwidth]{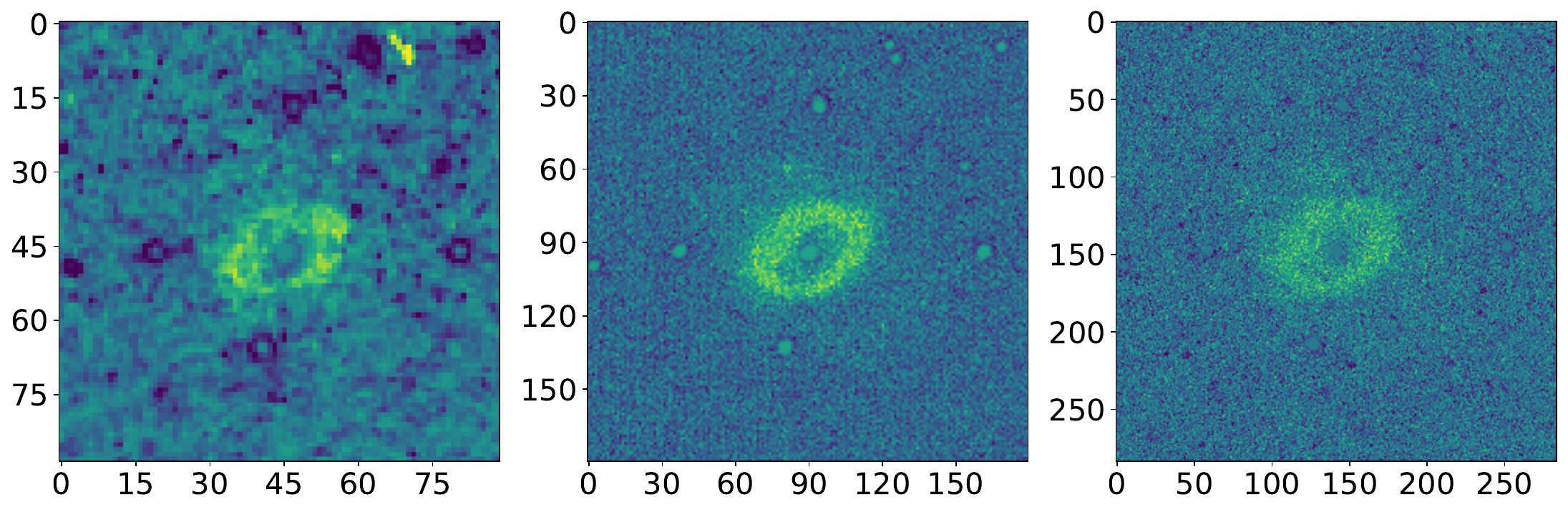}
\includegraphics[width=\columnwidth]{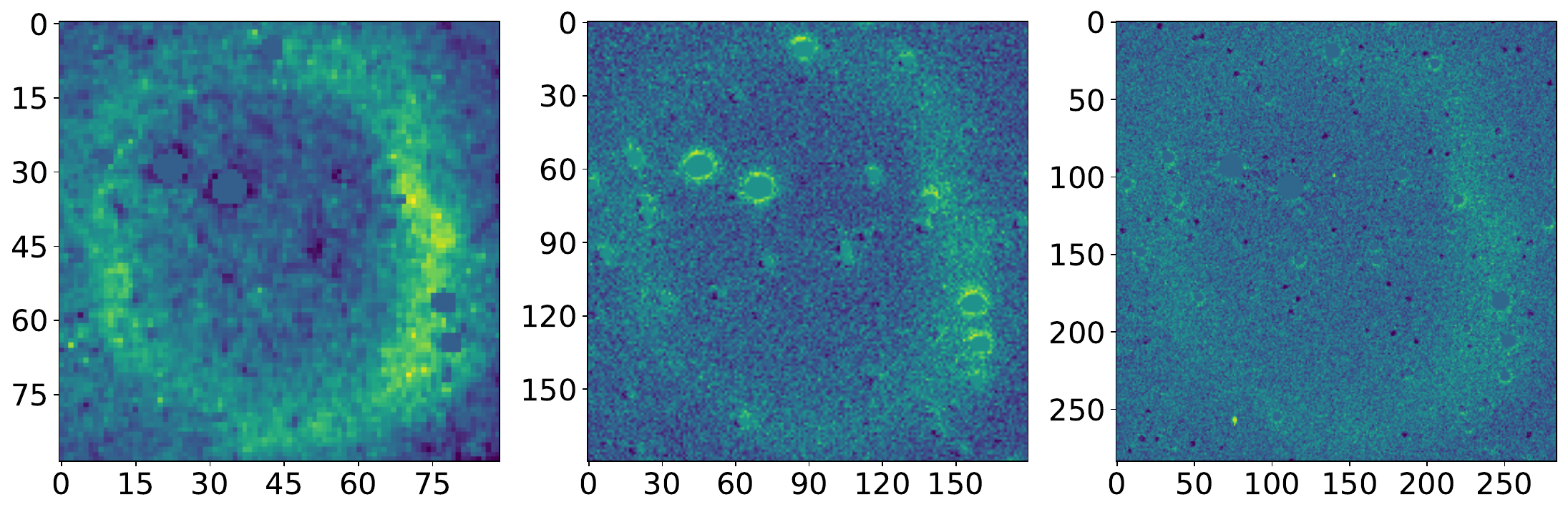}
\includegraphics[width=\columnwidth]{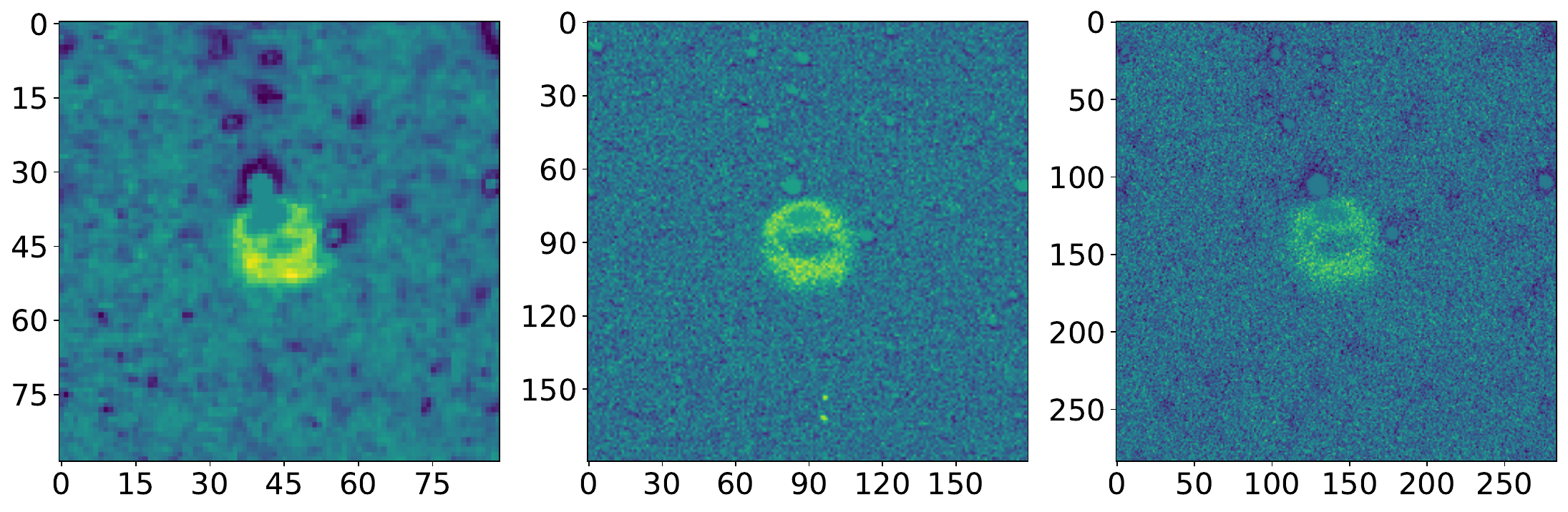}
\caption{The known PNe Abell 47, PHR J1843-0232, PHR J1843+0002 (HASH ID\# 352, 2469, 2478) in the SHS, IPHAS 
and VPHAS+ surveys respectively. The sensitivity of all 3 surveys for H$\alpha$ are similar while the resolution 
and so morphological detail are superior for the IPHAS and VPHAS+ surveys. The scale of these images is 60 arcsec on each side.}
\label{fig:surveys}
\end{figure}

The SuperCOSMOS H$\alpha$ survey (SHS) \citep{2005MNRAS.362..689P}, used the 1.2m UK Schmidt Telescope in Siding Spring 
Observatory, Australia, to undertake the first modern H$\alpha$ survey of the Southern Galactic plane between 1998 and 
2003. The survey covered over 4000~square degrees to $\pm$10 degrees in Galactic latitude at 1-2arcsec resolution
and to 5~Rayleigh sensitivity. It was the last great photographic survey ever undertaken. It used fine-grained Kodak 
Technical Pan film as a detector \citep{1999PASA...16..288P} that has a useful sensitivity peak at H$\alpha$ 
and was push-processed and hypersensitised in a way that provided 10\% DQE  (unprecedented for a photographic astronomical 
survey). It was undertaken with the world's largest single element optical interference filter \citep{1998PASA...15...33P}. 
All the films were digitised using the SuperCOSMOS microdensitometer \citet{2001MNRAS.326.1279H} and they have been 
carefully calibrated onto a Rayleigh Scale \citep{2014MNRAS.440.1080F}. Even today in terms of depth, resolution and 
coverage it remains competitive (see Figure \ref{fig:surveys}). In this work we only use it as a base reference for 
comparison with VPHAS+ and, in the overlap region, IPHAS.

\subsection{IPHAS}
\label{sec:iphas} 

The IPHAS survey used the Isaac Newton 2.5-meter Telescope (INT) on the island of La Palma, Canary Islands 
\citep{2005MNRAS.362..753D}. IPHAS covered $\sim$1800 square degrees sampling Northern Galactic latitudes $|b| < 
5^\circ$. It used a Wide Field Camera with a field of view of 0.3 square degrees covered by 4 CCDs with $2048\times 
4096$ pixels and with a resolution of up to 0.33 arcsec/pixel. The median seeing for survey-standard data is 
approximately 1.1 arcsec. The survey limiting magnitude can exceed r' $\approx$ 20 mag (10$\sigma$). The IPHAS data 
include observations with a H$\alpha$ narrow band filter and r' and i' bands of the Sloan Digital Sky Survey (SDSS). 
The H$\alpha$ band has a central wavelength of 6568 $\angstrom$ with a bandwidth of 95 $\angstrom$. The strategy of 
multiple repeat observations with slightly offset pointing field centers ensures that there are no omissions at the edges 
of the field of view and allows for better control of observational quality. The survey aimed to potentially increase 
the number of known emission line targets in the Northern hemisphere by an order of magnitude.

\subsection{VPHAS+}
\label{sec:vphas+} 

The VST Photometric H$\alpha$ Survey of the Southern Galactic Plane and bulge (VPHAS+) is very similar to the IPHAS survey, 
but focuses on the observations of the southern Galactic plane and bulge, and has a larger field of view and slightly 
higher resolution. The VPHAS+ project uses the VLT Survey Telescope (VST) for observations with a 2.6-meter aperture 
located in Cerro Paranal, Chile. The OmegaCAM camera has 32 CCDs with $2048\times 4096$ pixels, which allows for a 
field of view of 1 square degree and a resolution of up to 0.21 arcsec/pixel. The survey planned to observe approximately 
2000 square degrees of sky in a range of $\pm 5^\circ$ centered on the Galactic plane, with an extension to $\pm 
10^\circ$ near the bulge. The survey is divided into 2284 of $1^\circ\times 1^\circ$ fields, and $91.6\%$ of the goal 
has been completed. In addition to the H$\alpha$ narrow band, the survey also includes the u, g, r, and i broadband 
filters of SDSS. The aim of limiting magnitude of the survey is to reach at least 20 mag (5$\sigma$) in each band, with a typical 
seeing of approximately 1.0 arcsec and approximately 0.8-0.9 arcsec in selected dense star fields such as areas with less 
extinction in the Galactic bulge. For the observation strategy, each field of view is observed at three slightly different 
central positions to achieve full coverage and deeper depth.

\begin{table*}
\centering
\caption{Properties of H$\alpha$ Surveys VPHAS+, IPHAS and SHS.}
\label{tab:survey}
\resizebox{\textwidth}{!}{
\begin{tabular}{ccccccccccccccc} 
\hline
\multicolumn{6}{c}{Survey} & \multicolumn{4}{c}{Telescope} & \multicolumn{5}{c}{Camera} \\
\hline
Name & Hemisphere & Duration & Coverage & Area & Depth & Name & Aperture & Filter & Median Seeing & Name & CCD & Pixel & 
Resolution & Field of View \\
 & & & & deg$^2$ & mag & & m & & arcsec & & & & arcsec/pixel & deg$^2$ \\
\hline
\multirow{2}{*}{VPHAS+} & \multirow{2}{*}{Southern} & 2018.08 - now & |b|<5$^\circ$; 210$^\circ$<l<40$^\circ$ & \multirow{2}{*}
{$\sim 2000$} & >$\sim 20$ (All), & \multirow{2}{*}{VST} & \multirow{2}{*}{2.6} & H$\alpha$, & 1.0, $\sim$ 0.8-0.9 & \multirow{2}
{*}{OmegaCAM} & \multirow{2}{*}{32} & \multirow{2}{*}{2048$\times$4100} & \multirow{2}{*}{0.21} & \multirow{2}{*}{1} \\
 & & ($91.6\%$) & |b|<10$^\circ$; 350$^\circ$<l<10$^\circ$ & & 22 (g) & & & u, g, r, i & (dense field) & & & & & \\
\hline
\multirow{2}{*}{IPHAS} & \multirow{2}{*}{Northern} & \multirow{2}{*}{2003.08 - 2008} & \multirow{2}{*}{|b|<5$^\circ$; 
29$^\circ$<l<215$^\circ$} & \multirow{2}{*}{1800} & 21 (r), 20 (i), & \multirow{2}{*}{INT} & \multirow{2}{*}{2.5} & H$\alpha$,  & 
\multirow{2}{*}{1.2} & Wide Field & \multirow{2}{*}{4} & \multirow{2}{*}{2048$\times$4100} & \multirow{2}{*}{0.33} & \multirow{2}
{*}{0.3} \\
 & & & & & 20 (H$\alpha$)& & & r, i & & Camera & & & & \\
\hline
\multirow{2}{*}{SHS} & \multirow{2}{*}{Southern} & \multirow{2}{*}{1997.07 - 2003} & \multirow{2}{*}{|b|<10-13$^\circ$; 
Dec<2$^\circ$} & \multirow{2}{*}{4000} & \multirow{2}{*}{$\sim 20.5$} & \multirow{2}{*}{UKST} & \multirow{2}{*}{0.3} & H$\alpha$, 
& \multirow{2}{*}{1.0-2.0} & Super- & \multirow{2}{*}{1} & \multirow{2}{*}{2048$\times$2048} & \multirow{2}{*}{0.67} & \multirow{2}
{*}{5.5} \\
 & & & & & & & & short-r & & COSMOS & & & & \\
\hline
\end{tabular}}
\end{table*}

\section{Method}
\label{sec:method}
In this section we provide a detailed description of the methods and strategies we have adopted  based on a new deep 
learning technique to search for PNe candidates in narrow-band wide-field H$\alpha$ survey imagery. 
In Section \ref{sec:model}, we briefly introduce the model, and in Section 
\ref{sec:dataset}, we  further describe the method of constructing the dataset and the different data processing 
methods used for the IPHAS and VPHAS+ surveys.

\subsection{Model}
\label{sec:model}
\begin{figure}
\includegraphics[width=\columnwidth]{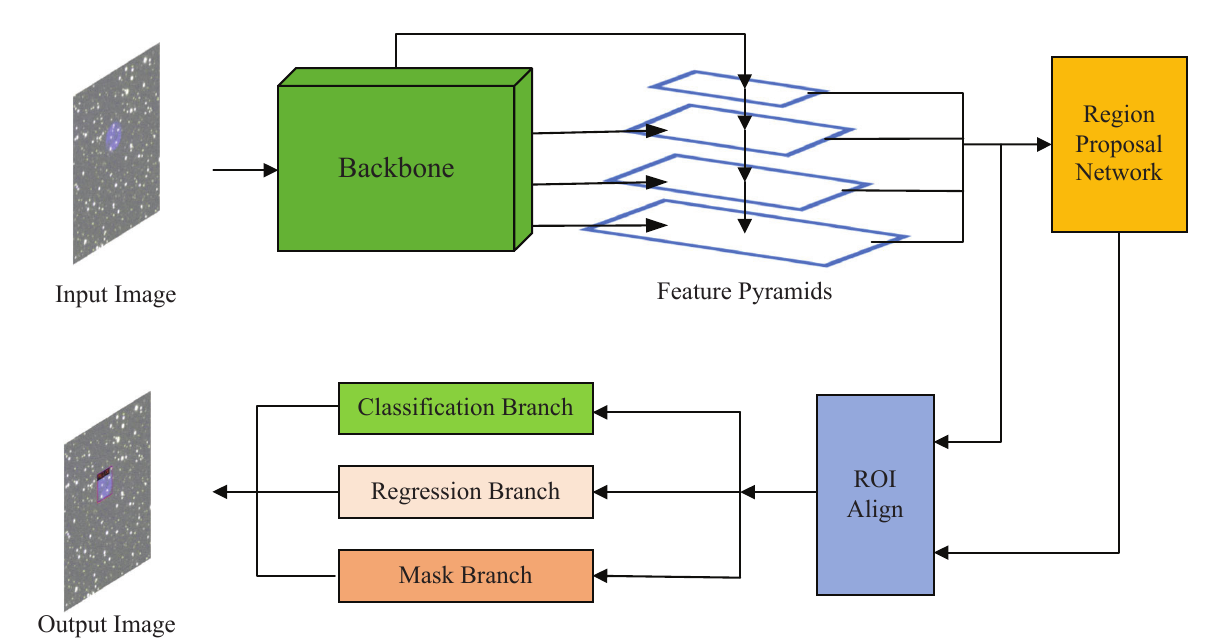}
\caption{The structure of Swin-Transformer model in this paper.}
\label{fig:structure}
\end{figure}

The overall structure of the model is shown in the Figure \ref{fig:structure}. When an image is input to the model, it 
first undergoes feature extraction through a feature extraction network and uses so-called pyramid technology 
\citep{lin2017feature} to output multi-scale feature maps at different levels. These multi-scale features are first 
detected through an Region Proposal Network (RPN) module e.g. \citep{ren2015faster}, which outputs a series of bounding 
boxes that may contain a target of interest. The ROI Align module \citep{he2017mask} is then used to extract the corresponding 
feature map parts of these bounding boxes. Finally, these extracted feature maps that may contain a target are sent 
to subsequent classification, position regression, and mask segmentation modules for further accurate classification 
and precise positioning, resulting in the final prediction results.

The model structure we used is similar to Mask R-CNN \citep{he2017mask}, which simultaneously predicts object classes, 
bounding boxes, and pixel-wise masks, making it a two-stage object detection architecture that typically has better 
detection accuracy than one-stage object detection networks such as YOLO \citep{2015arXiv150602640R}.

In the feature extraction module, compared to convolutional neural networks (CNNs), transformer based neural networks 
have no limitation on receptive fields and can model the relationships between any input features across any scale 
and distance, at the cost of larger GPU memory and longer training or inference time. Therefore, they have more powerful feature extraction capabilities and are more suitable for extended 
astronomical objects with complex structures hidden behind dust and/or in dense star fields such as PNe. 
Therefore, we chose `transformer' as the feature extraction module, whose core is the 
attention mechanism where feature extraction also depends on attention calculation.

\begin{figure}
\includegraphics[width=\columnwidth]{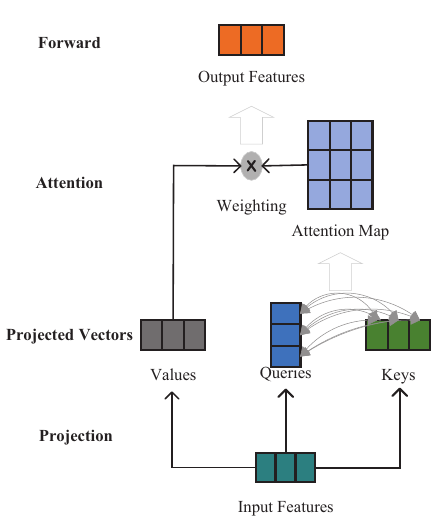}
\caption{The procedure for calculation of attentions by the transformer.}
\label{fig:attention}
\end{figure}

\begin{figure*}
\includegraphics[width=0.6\textwidth]{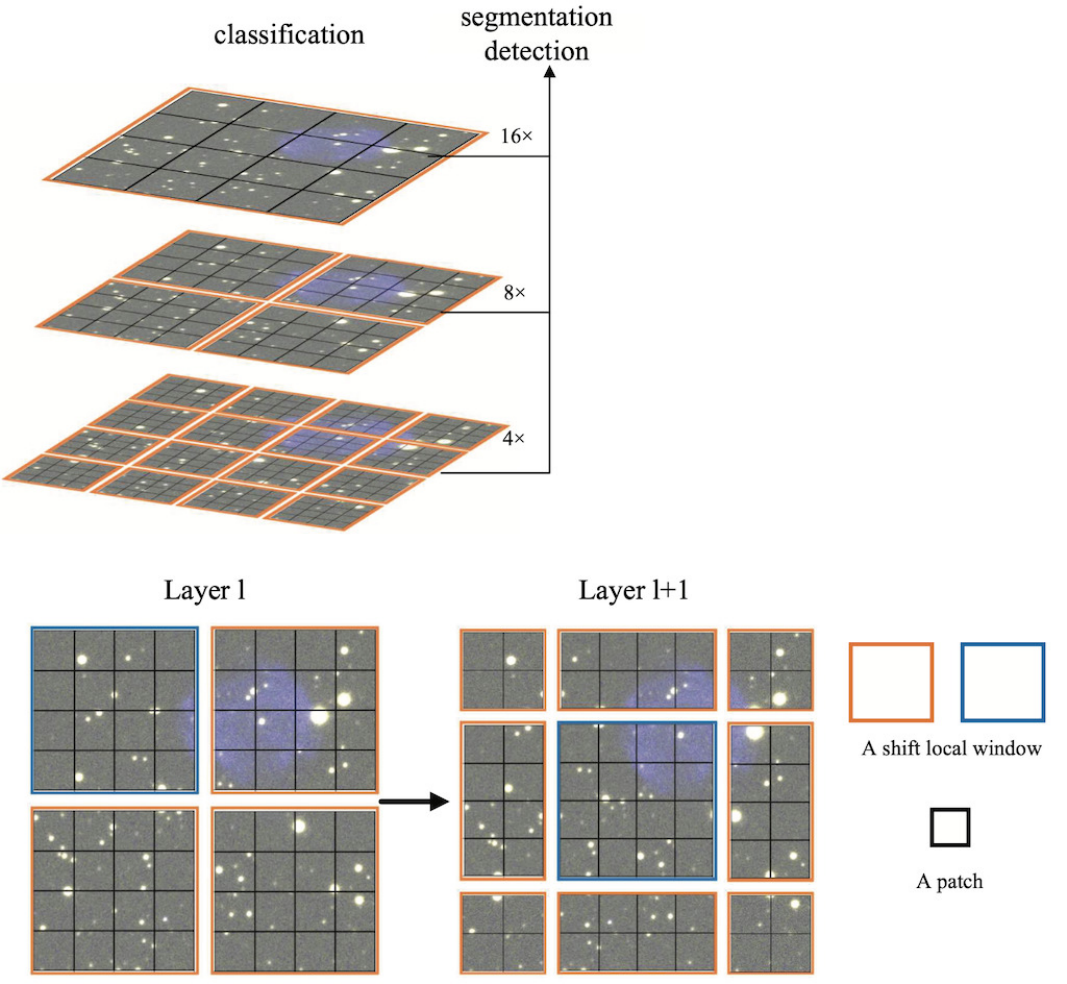}
\caption{This figure demonstrates the multi-level features and how the shifted local window captures 
various image components for target detection purposes.}
\label{fig:moving_window}
\end{figure*}

The calculation process is shown in the Figure \ref{fig:attention}. When a feature vector is processed, 
it is first projected onto different directions and positions in the feature space to transform 
into three vectors: Value, Query, and Key. This is similar to the hidden state in Recurrent Neural 
Network. Value, Query, and Key respectively contain different distribution 
information of the feature vector in the feature space and are used to calculate the relationship between the current 
feature vector and other feature vectors. For each feature vector, its Query is dot-producted with the Key of all feature 
vectors to obtain the correlation between the current feature vector and all feature vectors. For each feature vector, we 
can calculate the relationship with other feature vectors to obtain an attention map, where the value of each pixel in the 
map represents the correlation between these corresponding features. The higher the value of the correlation, the closer the 
relationship between these two features, and the higher the model's attention. This calculation method is based 
on something akin to 'intuitive feeling': i.e. if two feature vectors are closer, their similarity is higher, 
so they should have a greater weight. The final attention map can be regarded as an attention distribution and is normalised through a so-called softmax layer, which uses the 
softmax function to calculate the distribution that the input example belongs to a specific function \citep{rumelhart1986learning}. 
Thus, the attention map between different vectors is transformed into a weight map between different features, and the 
weight map output by the softmax layer is dot-producted with the Value of all corresponding feature vectors to obtain the 
final output. Through this structure, the attention layer naturally obtains the connection between features, regardless of 
their distance, breaking through the limitation of the receptive field of convolutional kernels in CNNs.

\begin{figure*}
\includegraphics[width=0.6\textwidth]{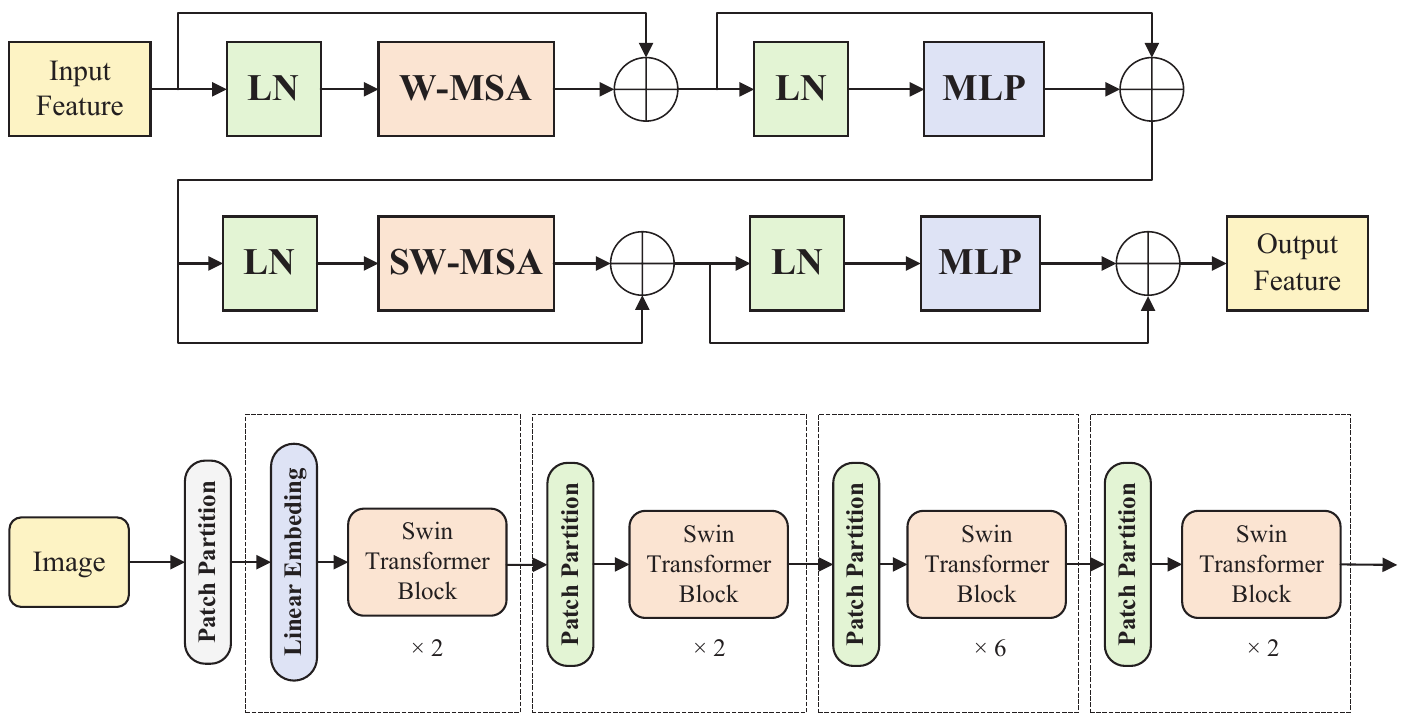}
\caption{The upper panel of this figure displays the feature extraction module, where the input features undergo processing 
through the 'W-MSA' and 'SW-MSA' components to generate the output features. The complete structure of the Swin-Transformer 
feature extraction module is presented in the lower panel of this figure.}
\label{fig:swin_structure}
\end{figure*}

However, the calculation of attention in the transformer is a dense operation that involves interactions between each pair 
of inputs, which means that the complexity of the transformer model increases by the square of the size of the input 
image. Since astronomical images can be large, this is computationally unacceptable. Therefore, we improved the transformer
structure by using the Swin-Transformer structure, which retains the powerful feature extraction capabilities of the 
transformer while effectively solving the problem of high model complexity \citep{jia2023deep}. Its core mainly 
includes two parts: the shifted windows and multi-level feature, as shown in Figure \ref{fig:moving_window}.

Firstly, we elaborate on the design of the shifted windows that we employ. In Figure \ref{fig:moving_window}, we can 
see that unlike the attention calculation of the transformer structure, which directly interacts between all pixels of 
the input image, the Swin-Transformer first divides the input image into numerous small windows and then performs the 
attention calculation between different patches individually within each small window. This greatly reduces the 
complexity of the model, reducing the transformer's $\rm N^2$ model complexity to linear complexity. At the same time, 
we noticed that if only window partitioning is performed, there will be a lack of information exchange between 
different windows, making it difficult to model larger-scale features that span across several windows. 
Therefore, we further refined the shifted window method as shown in 
Figure \ref{fig:moving_window}. The rule windows partitioned at layer l is shifted when it passes to the next layer 
l+1, generating new windows. The attention calculation is performed within the new windows. As a result, 
information exchange between the windows of layer l is achieved within the new windows of layer l+1, 
helping the model to better extract features.

Secondly, we briefly describe the multi-level features of our model. 
From Figure \ref{fig:moving_window}, we can see that the Swin-Transformer performs 
feature extraction on the input image at different levels. The features at different levels are 
stacked together to form a feature pyramid that is input to the subsequent modules. This helps 
the model obtain more comprehensive multi-scale feature information and achieve more accurate 
detection results \citep{lin2017feature}.

The overall structure of the Swin-Transformer feature extraction module is shown in Figure \ref{fig:swin_structure}. 
Here, when performing feature extraction on the input image, the image is first 
divided into patches and windows using so-called `Patch Partition', and then 
transformed into a sequence form that the transformer can 
accept using Linear Embedding, see \citet{chen2011locally}. The input sequence undergoes several 
consecutive Swin-Transformer blocks to extract features, and undergoes down-sampling via Patch 
Merging to increase the receptive field of the model and extract multi-scale features at different levels. 
The internal structure of two consecutive Swin-Transformer blocks is shown in Figure 
\ref{fig:swin_structure}. In the first Swin-Transformer block, attention calculation is performed within 
the regularly partitioned windows, while in the second Swin-Transformer block, attention calculation 
is performed within the shifted windows.

In addition, it is worth noting that since attention calculation itself ignores the positional 
relationship between features, position encoding is needed before inputting features into the 
transformer. Here, we use a technique called relative position encoding, e.g. \citet{DBLP:journals/corr/abs-2107-14222}. 
Moreover, the loss function used in our model is consistent with Mask R-CNN \citep{he2017mask}.

\begin{figure*}
\includegraphics[width=0.8\textwidth]{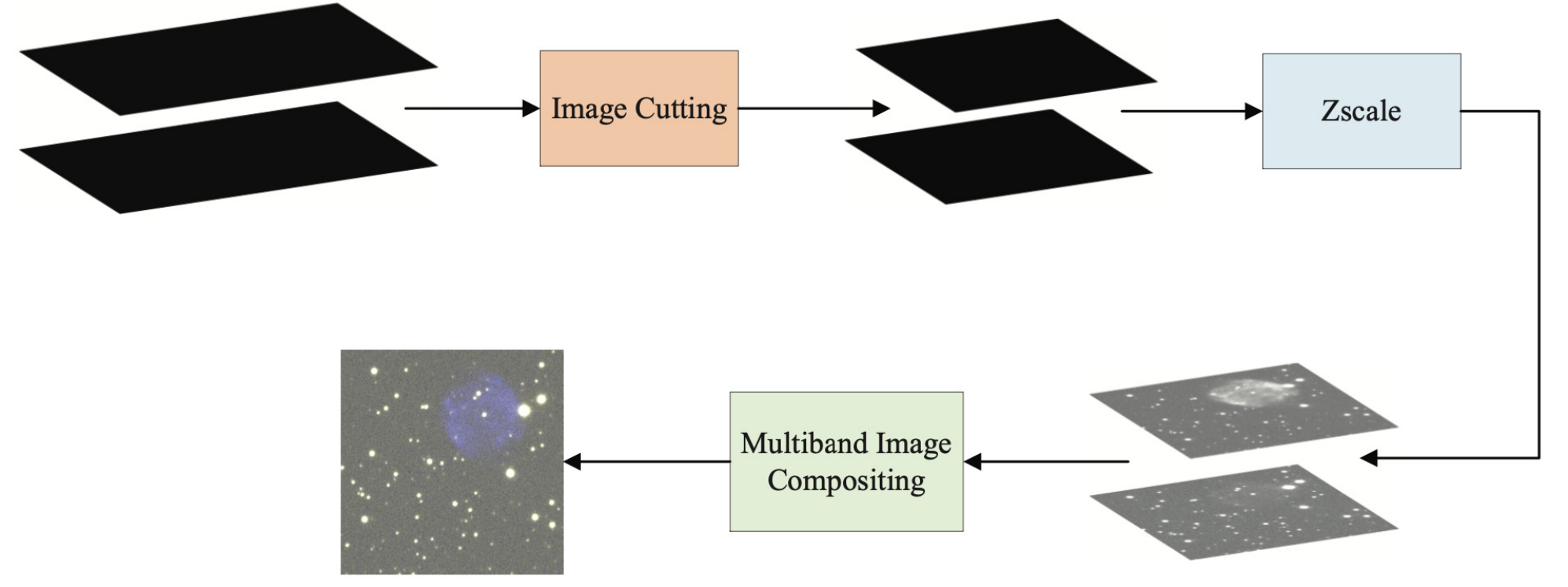}
\caption{This figure illustrates the complete pre-process procedure of constructing the dataset. 
As depicted, the preprocessing stage encompasses three key components: image cropping for subsequent 
data processing, grayscale transformation to enhance signal-to-noise ratio, and the integration of 
multi-band image data to create a merged, colored image.}
\label{fig:dataset}
\end{figure*}

\subsection{Dataset}
\label{sec:dataset}

\begin{figure}
\includegraphics[width=0.5\columnwidth]{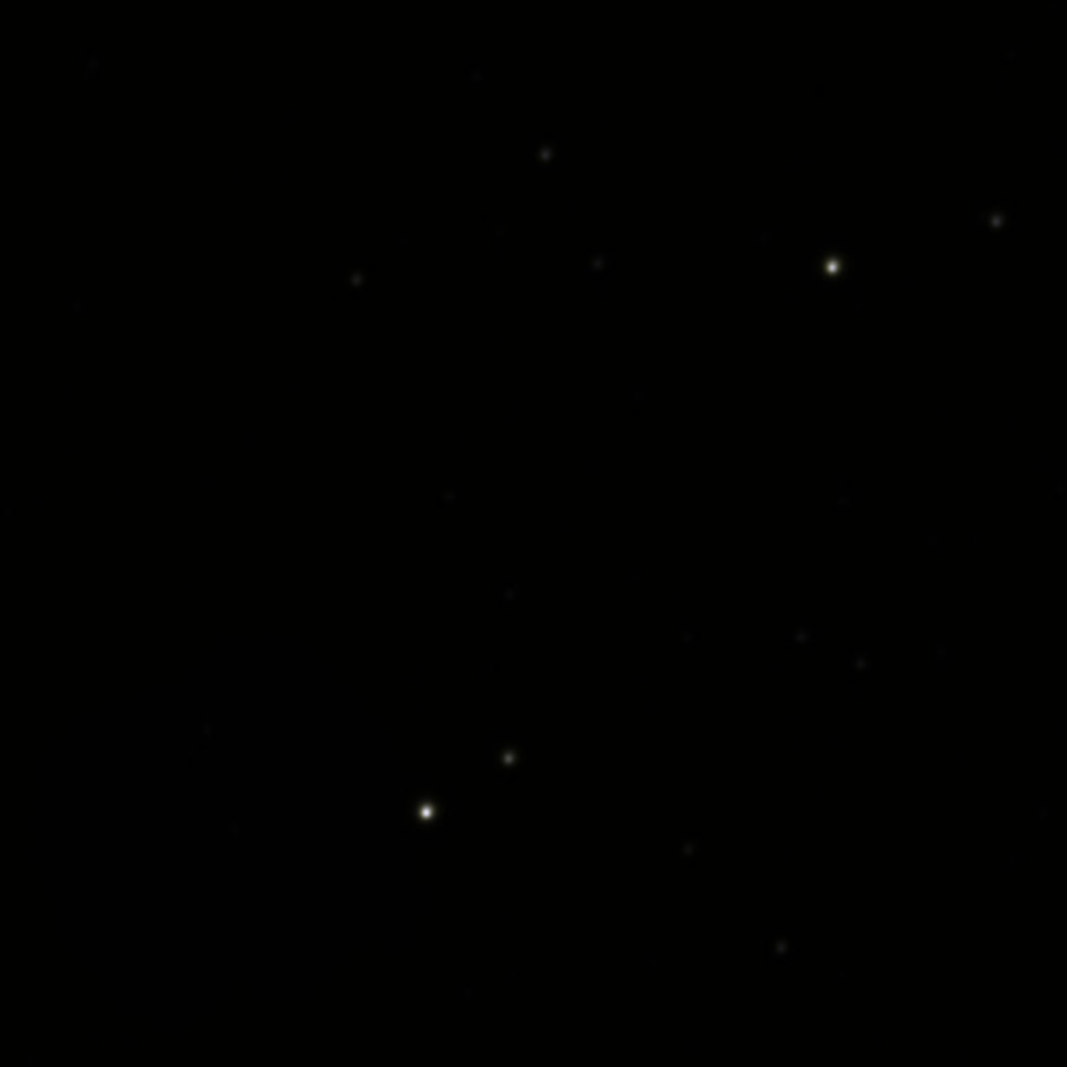}
\includegraphics[width=0.5\columnwidth]{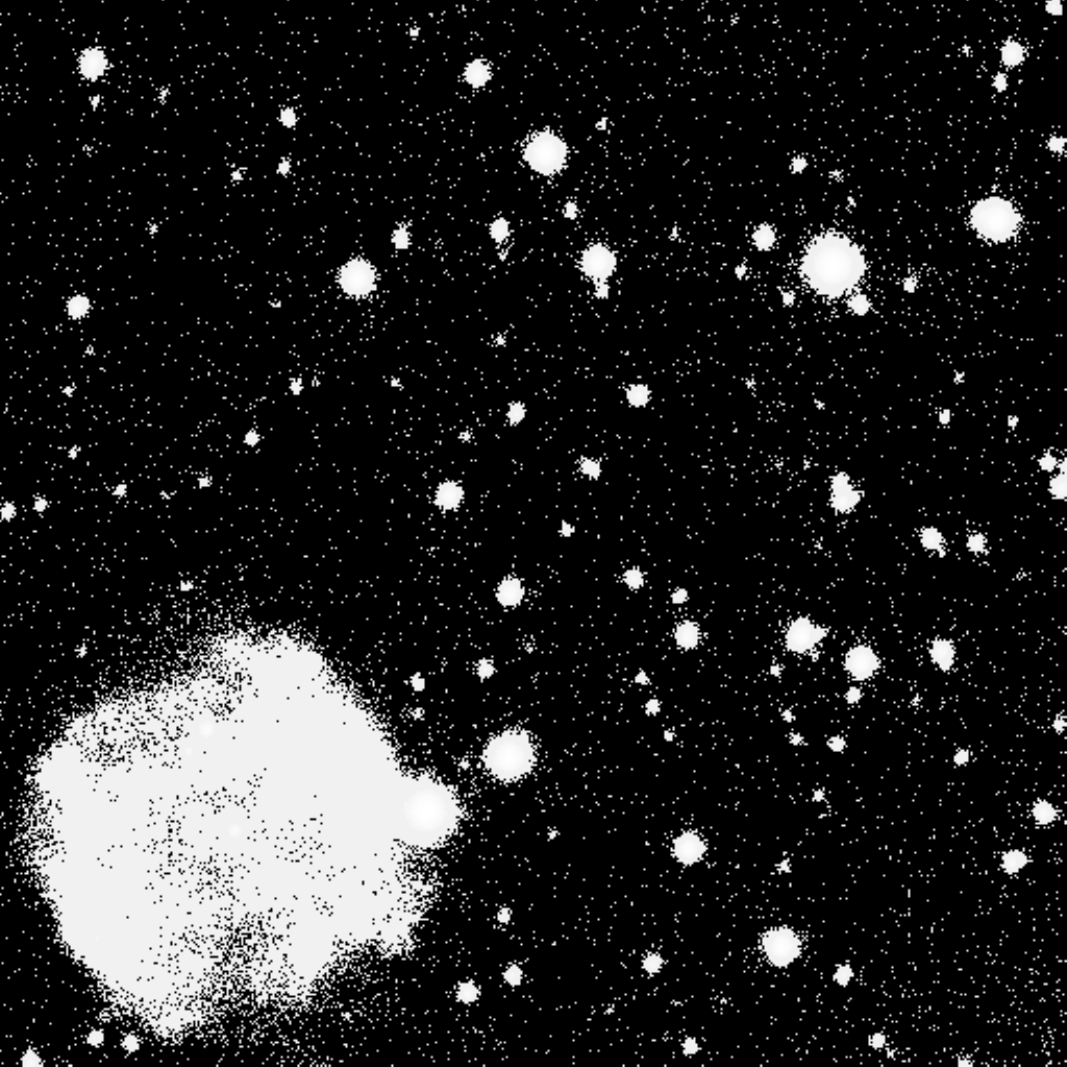}
\includegraphics[width=0.5\columnwidth]{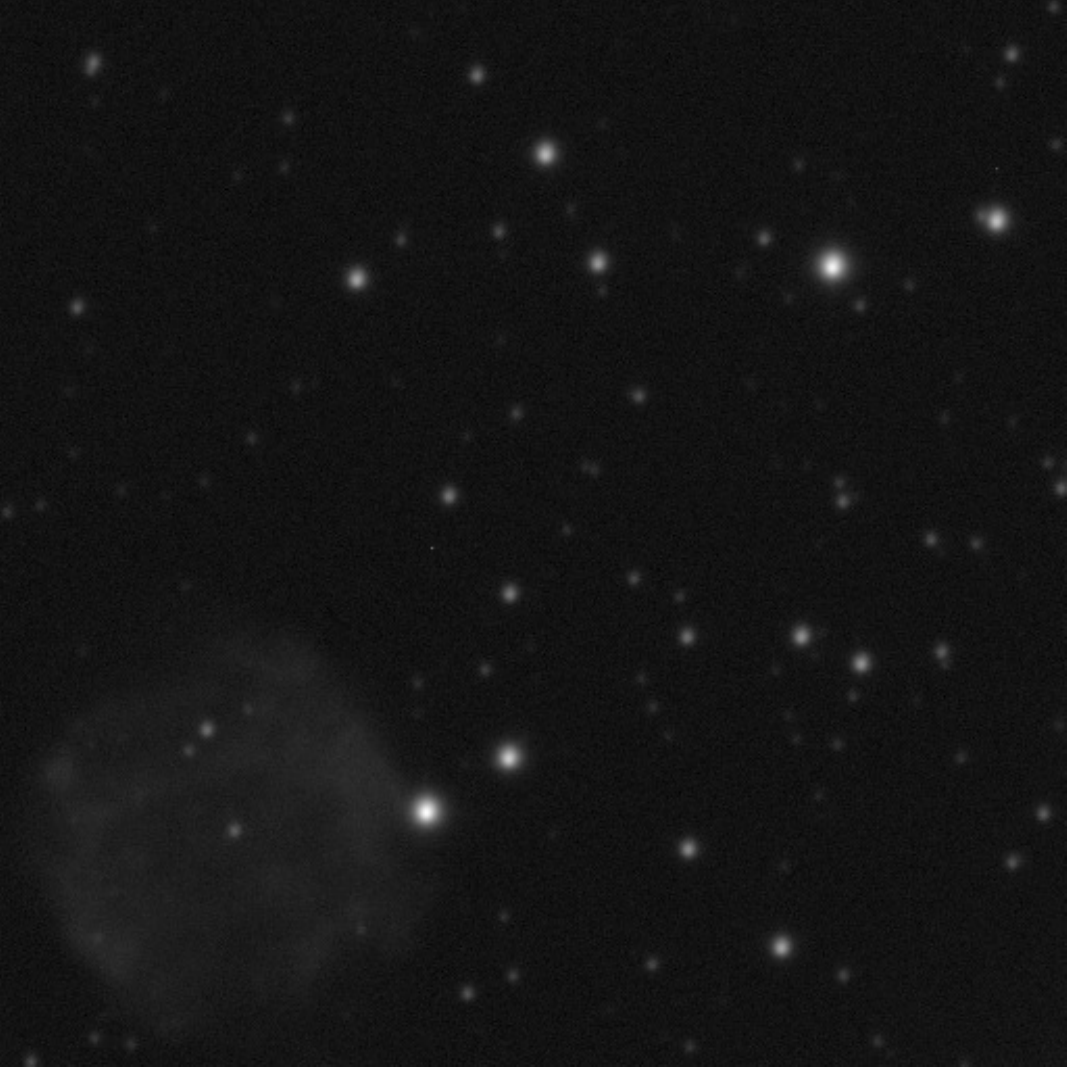}
\includegraphics[width=0.5\columnwidth]{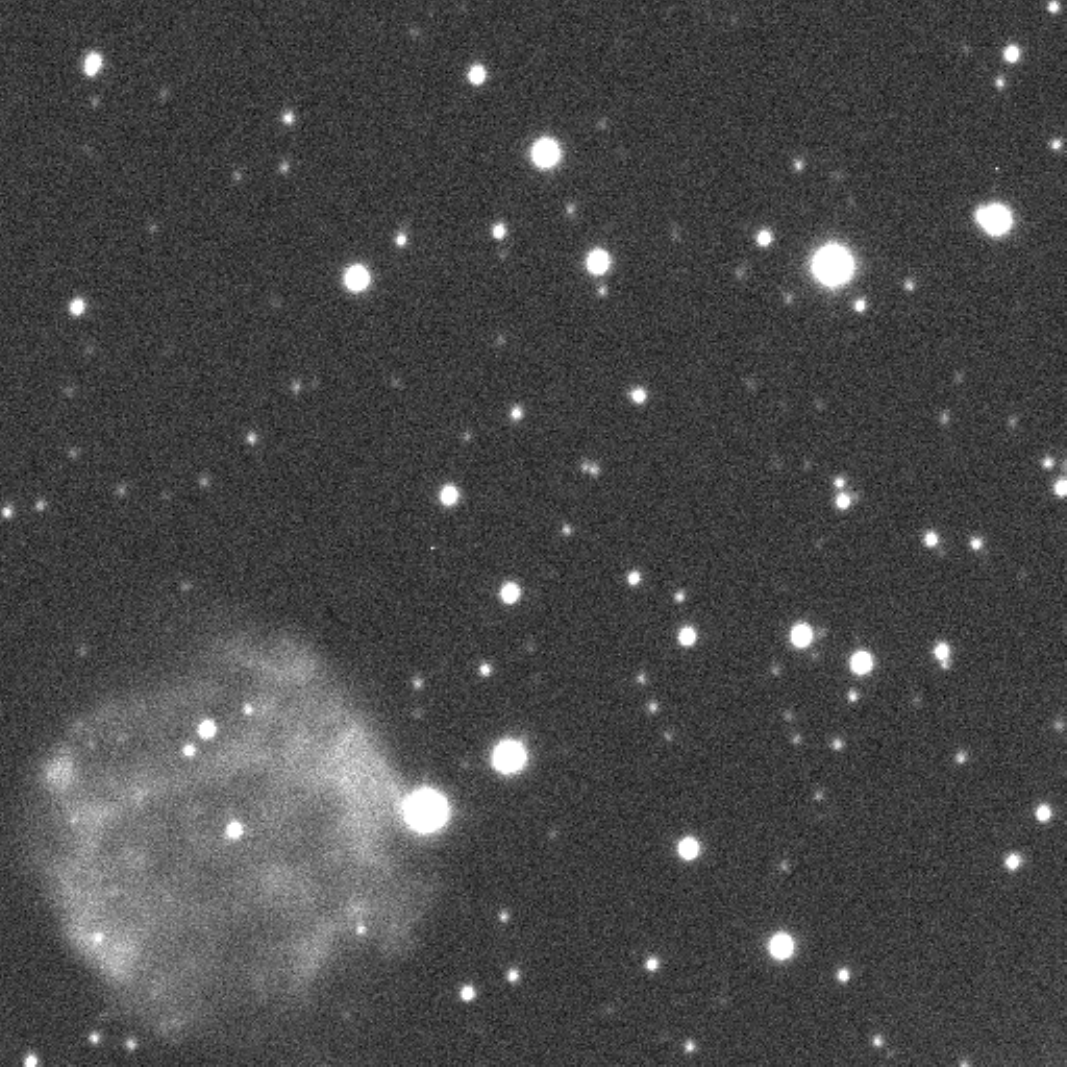}
\caption{This figure displays images subjected to various grayscale transformations. The top-left panel features 
the original image, while the top-right panel showcases the image after sinh transformation. In the bottom panel, 
we observe the image after log transformation, and in the bottom-right figure, we show the image after zscale 
transformation. As demonstrated in this figure, the z-scale transformation effectively enhances the signal-to-noise 
ratio of target objects.}
\label{fig:transform}
\end{figure}

As the detection model we build is a data-driven algorithm, we need to further construct appropriate data 
sets to train and validate the model's performance. Due to the stable and 
essential uniform quality of IPHAS survey data, e.g. \citep{2021A&A...655A..49G} and the existing systematic 
visual search already undertaken to find PNe , a large number of high-quality PNe have already been found for this purpose, 
e.g.  \citep{2014MNRAS.443.3388S}. Therefore, using IPHAS images and catalog data, we can easily construct 
a dataset and corresponding labels that can be used for both model training and model evaluation. 
We have processed the IPHAS data to better adapt it to our model. Firstly, as the size of original IPHAS
images are $4096 \times 2048$ 0.33 arcsec pixels and considering the 
model complexity and hardware limitations, we cut these images into overlapping $512 \times 512$ 
small images. Secondly, we used a series of different gray-scale transformation techniques for image enhancement 
to improve model performance, as shown in the Figure \ref{fig:transform}. It can be seen that the 
image contrast has greatly improved after the gray-scale transformation, making it easier to identify candidate 
PNe. This greatly helps accelerate model convergence and improve detection accuracy. 
After several tests, we finally selected the Zscale gray-scale transformation \citep{2000ascl.soft03002S} as the most useful.

Most PNe are usually obvious in uncrowded, uncomplicated regions of the narrow-band H$\alpha$ survey 
observations. In IPHAS, the observation time of the H$\alpha$ band was adjusted to achieve 
consistency of image depth between it and the accompanying `r' broad-band image. 
Therefore, compared with the r band, the signal 
of emission objects in the H$\alpha$ band will be enhanced, while the signals of other continuum 
astronomical sources do not change significantly. Therefore, to highlight the image features of 
the PNe, we combine the images of these two bands into a multi-channel image. In this process, 
due to the existence of certain pixel offsets between the images of the two bands, we use 
Reproject\footnote{ https://reproject.readthedocs.io/en/stable/} to achieve the pixel-level precision 
alignment of the images of the two bands needed. Finally, we put the image data of the H$\alpha$ band 
into the Blue channel, and the image data of the r band into the Red and Green channels and merge them into a PNG image. Although some information may be lost, the gray-scale distribution of the image 
is more uniform, which is beneficial to the convergence of the model.

The whole process to construct the dataset is shown in Figure \ref{fig:dataset}. PNe 
exhibit a wide variety of different angular sizes from a few arcseconds up to arcminutes in the survey imagery. 
In order to allow the model to more accurately 
locate the target PNe, we use the maximum diameter of each nebula as the side length of the boundary box in 
its label. With this method, we can better constrain the model's localization of the PNe, thereby enabling the model to 
more accurately identify and analyze PNe of different sizes. After the above steps, we have built a training set 
and a validation set using IPHAS data. The IPHAS training set consists of 1137 cropped images with size of 
$512 \times 512$ pixels ($2.82 \times 2.82$ arcminutes) and the validation set contains 454 cropped images with 
size of $512 \times 512$, both sets of which contain known PNe.

\subsection{The VPHAS+ dataset}
\begin{figure}
\vspace*{-0.4in}
\hspace*{-0.3in}
\vspace*{-0.3in}
\includegraphics[width=1.2\columnwidth]{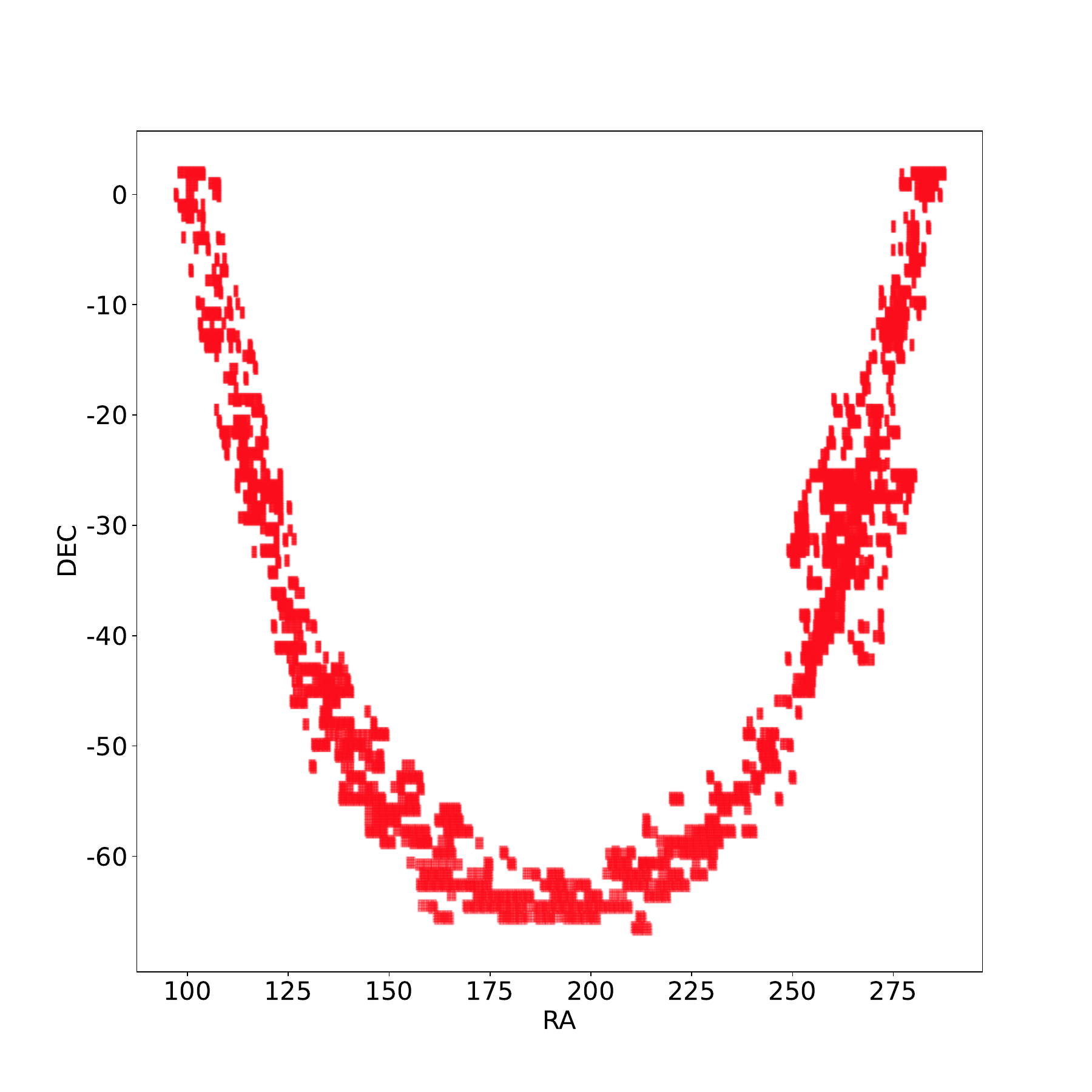}
\caption{This figure shows the distribution of the selected 979 our of 2284 VPHAS+ survey fields that were 
searched and projected in J2000 equatorial celestial coordinates of RA and DEC.}
\label{fig:field}
\end{figure}

Our key target H$\alpha$ survey using the above techniques for PNe candidate identification is the newer, 
Southern VPHAS+ Galactic plane survey \citep{2014MNRAS.440.2036D}. 
For the available VPHAS+ dataset, we performed the same cropping and gray-scale transformation as for IPHAS and also 
converted them into PNG images in the same way. However, there is a difference in image alignment because the image data 
of the two bands in VPHAS+ are not one-to-one as for IPHAS. Hence, it is not possible to directly find the 
corresponding images of the two bands. Therefore, we first selected the FITS images of the H$\alpha$ band and then used 
these images to match the corresponding images of the r band with pixel differences within 10. These matched data were 
combined into a multi-channel image to obtain the final PNG image. We processed $\sim$90,000 VPHAS+ images covering 
an area of 979 of the 2284 square degrees of the VPHAS+ survey. This is shown in Figure \ref{fig:field}. As a result 
we obtained $\sim$4.5 million PNG images of size $512 \times 512$ pixels (equivalent to 107 $\times$ 107 arcseconds) 
for the model's inference prediction and to search for new PNe targets.

\section{Results}
\label{sec:results}
The model we used was implemented based on the PyTorch framework \citep{paszke2019pytorch} and optimized with 
AdamW, a stochastic gradient descent method that is based on adaptive estimation of first-order and second-order moments
\citep{2017arXiv171105101L}. To ensure faster and better convergence of the model, we loaded pre-trained weights from the 
\citet{2021arXiv210314030L} to initialize the model. We deployed the model on a 
work station equipped with Nvidia RTX 3090 Ti GPU cards for training, validation, and candidate searching.

\subsection{Performance Evaluation Criterion}
\label{sec:Performance Evaluation Criterion}

We used the training and validation sets constructed from the IPHAS observation data in Section \ref{sec:dataset} to train 
and validate the model for use with the VPHAS+ survey data. To better evaluate the performance of the model, 
we adopted the following performance metrics.

The IOU (Intersection over Union - a metric used to evaluate Deep Learning algorithms by estimating how well a 
predicted mask or bounding box matches the ground truth data) is an important performance evaluation criterion 
in object detection. It is defined as the ratio between the overlap 
area and the union area of the bounding box from the detection results and the labels as defined in equation \ref{equation1}. 
The IOU quantifies the overlap between the detection bounding box and the real target bounding box, ranging from 0 to 1. In 
general, when the IOU is larger than a set threshold, the detection is considered to have correctly detected the target.
\begin{equation}
IOU = \frac{Intersection Area}{Union Area}
\label{equation1}
\end{equation}

According to the IOU evaluation criterion, we counted all the predicted results, and further quantitatively 
measured the overall detection performance of the model using the precision rate and the recall rate.
The precision rate is the percentage of true, positive detection results to all detection results and the recall 
rate is the percentage of true positive detection results to all targets. The precision rate and the recall rate 
are defined by the equation below: 
\ref{equation2}. 
\begin{equation}
\begin{split}
Precision = \frac{TP}{TP+FP}\\
Recall = \frac{TP}{TP+FN}
\label{equation2}
\end{split}
\end{equation}

Where TP is the number of targets correctly detected, FP is the number of negative samples incorrectly labeled as 
positive and FN is the number of true targets that have not been detected. A higher precision rate indicates a higher 
accuracy of the model. A higher recall rate indicates that the model has a strong ability to detect all real targets.

The precision and recall obtained by different IOU thresholds are often different. By calculating the precision and 
recall rates under different IOU thresholds, the precision-recall curve (PR curve) can be drawn. In the curve, 
the horizontal coordinate of a point is the recall rate, and the vertical coordinate is the precision rate. The area 
below the curve is the Average Precision (AP). The larger the area, the higher the AP value, the better the model performance.

\subsection{Results from IPHAS}
\label{sec:results in the IPHAS}
\begin{figure}
\includegraphics[width=\columnwidth]{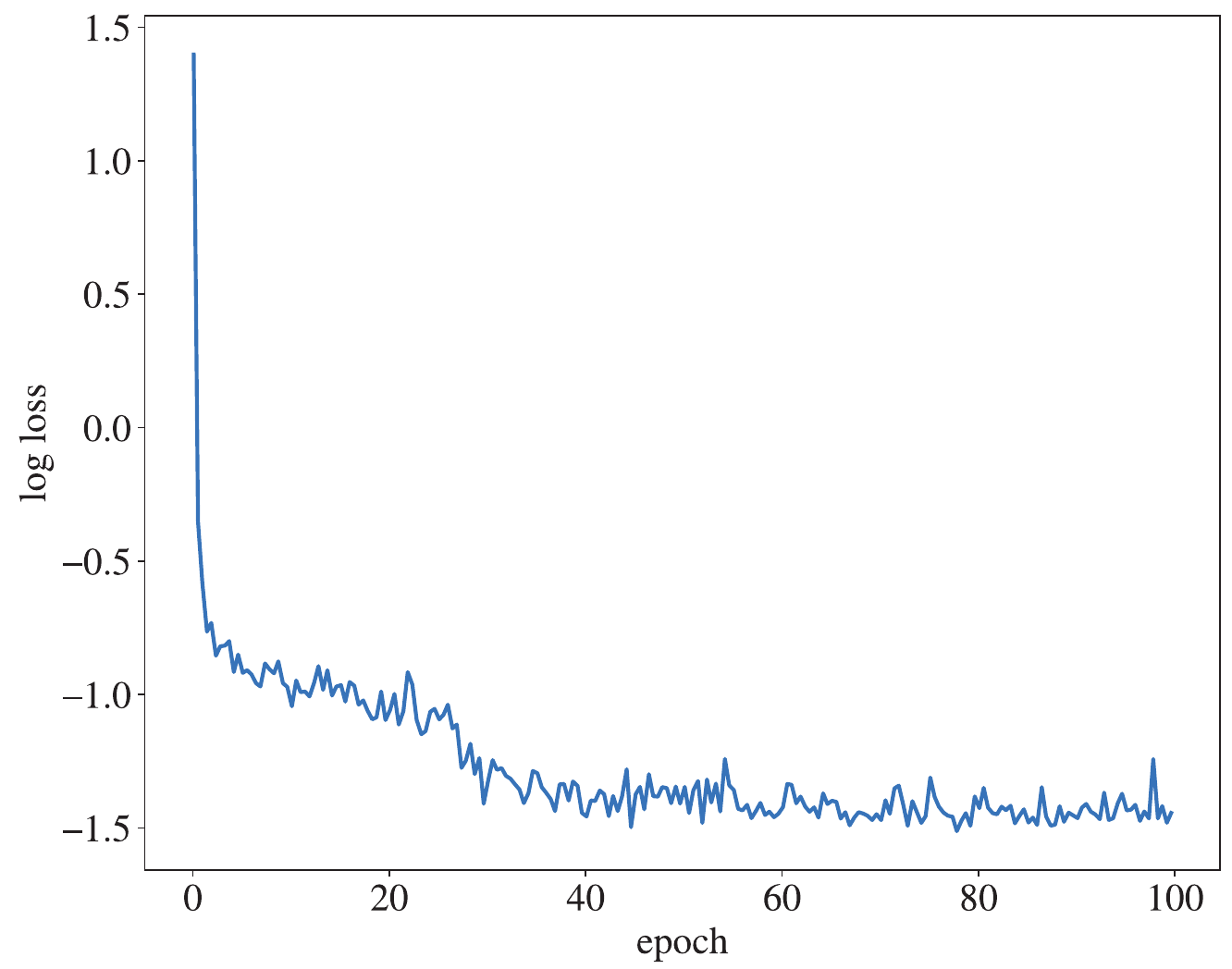}
\caption{The value of the loss is going down with the training.
As depicted, the neural network begins to converge at approximately 40 epochs, signaling our decision to 
stop training after this point.}
\label{fig:loss}
\end{figure}

\begin{figure}
\includegraphics[width=\columnwidth]{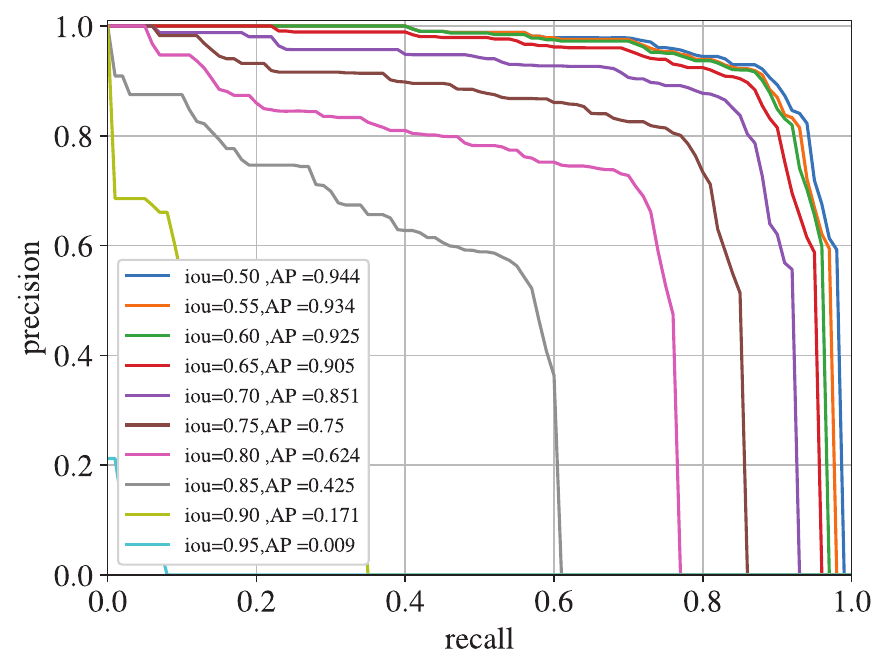}
\caption{The model's performance on the validation set is assessed with varying IOU thresholds, as demonstrated 
in this figure. When we set the IOU to 0.5, the AP can reach approximately 0.944. This suggests that, 
in our detection algorithm, we have the option to prioritize higher detection accuracy and recall rate at the 
expense of location accuracy. However, employing a lower iou value comes with the 
trade-off of requiring more human vetting near the positions identified by the detection algorithm.}
\label{fig:pr}
\end{figure}

\begin{figure}
\includegraphics[width=\columnwidth]{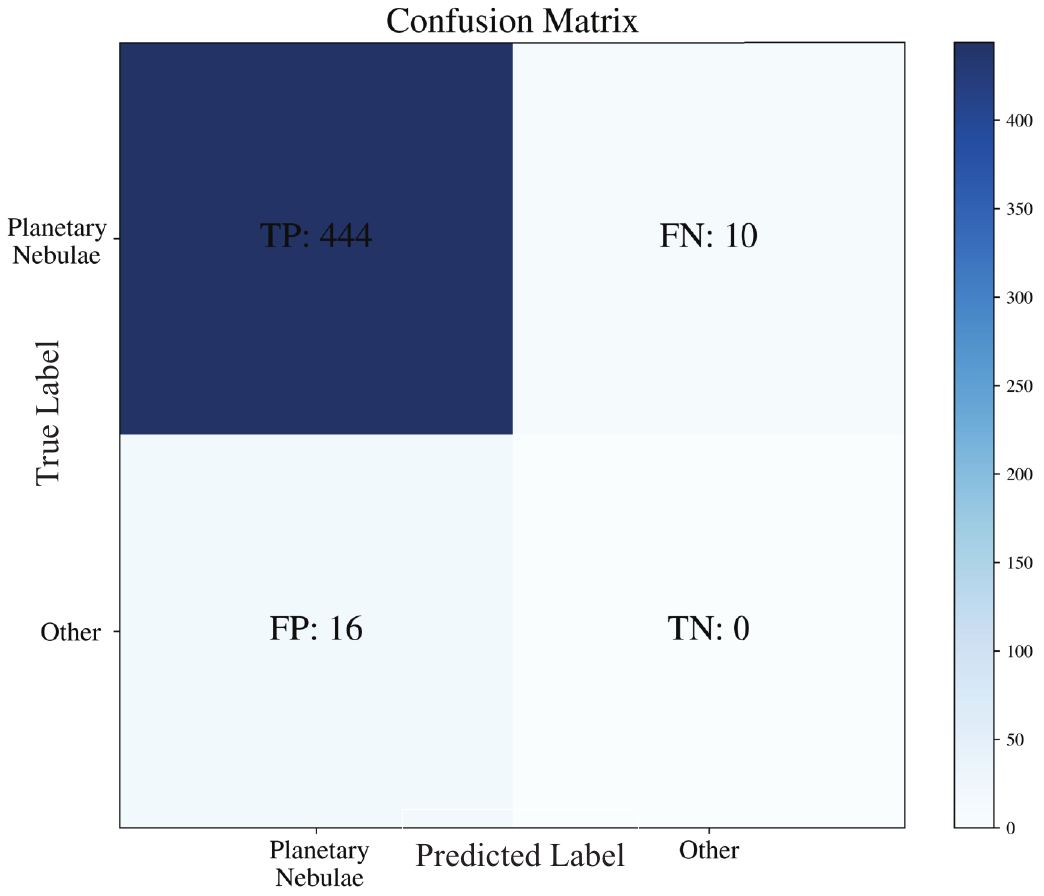}
\caption{The confusion matrix for the model on the validation set is presented in this figure. It's evident that our algorithm demonstrates high accuracy, correctly detecting the majority of targets with minimal occurrences of false positives and false negatives.}
\label{fig:confusion}
\end{figure}

The model was trained with 100 epochs (iterations), which took 10 hours computation on our computer server with one RTX 
3090 Ti GPU. The entire training process is shown in Figure \ref{fig:loss}, which shows that our model 
convergences after around 40 epochs. The performance of the model on the validation set is shown as the 
confusion matrix in Figure \ref{fig:pr}. Our model achieves ``Average Precision'' (AP) of 0.944, when the IOU 
is set as 0.5 which we set the IOU threshold. We have further made statistics on the prediction results 
when the IOU threshold was 0.5, as shown in the Figure\ref{fig:confusion}. As can be seen, out 454 known
targets in the IPHAS validation set, we successfully identified 444 and had 16 false positives. 
This achieved a recall rate of $97.8\%$ and a precision rate of $96.5\%$.

\begin{figure}
\includegraphics[width=0.5\columnwidth]{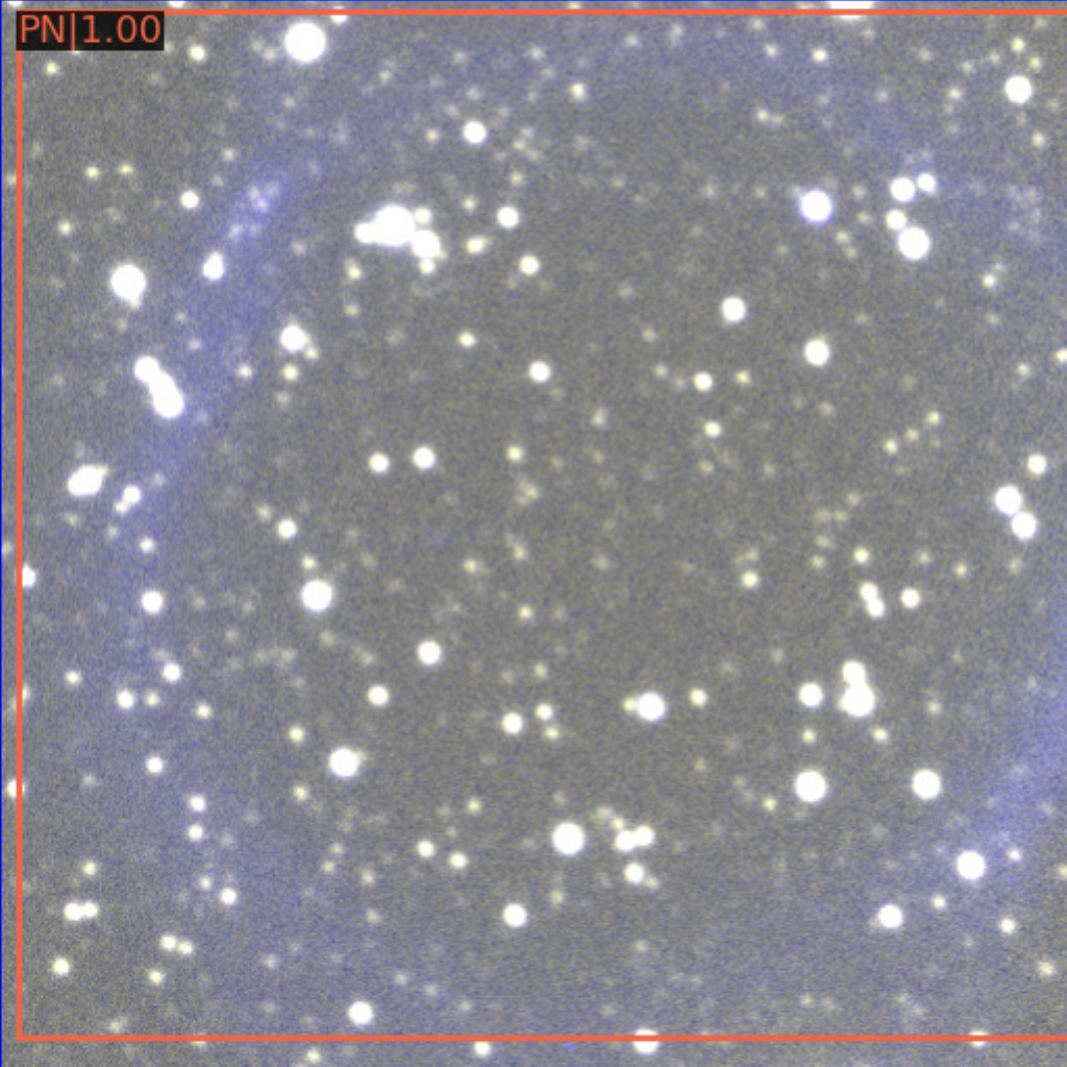}
\includegraphics[width=0.5\columnwidth]{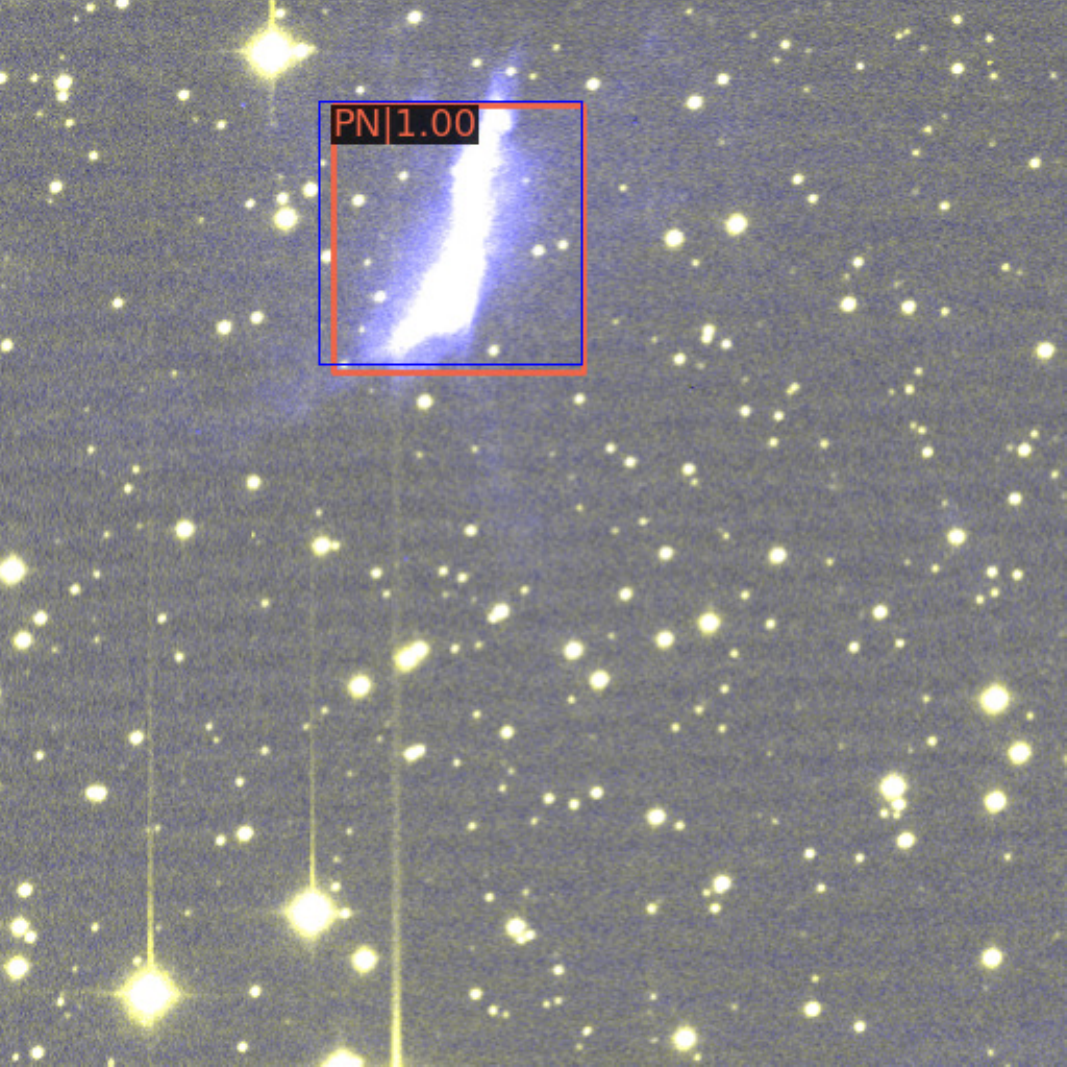}
\includegraphics[width=0.5\columnwidth]{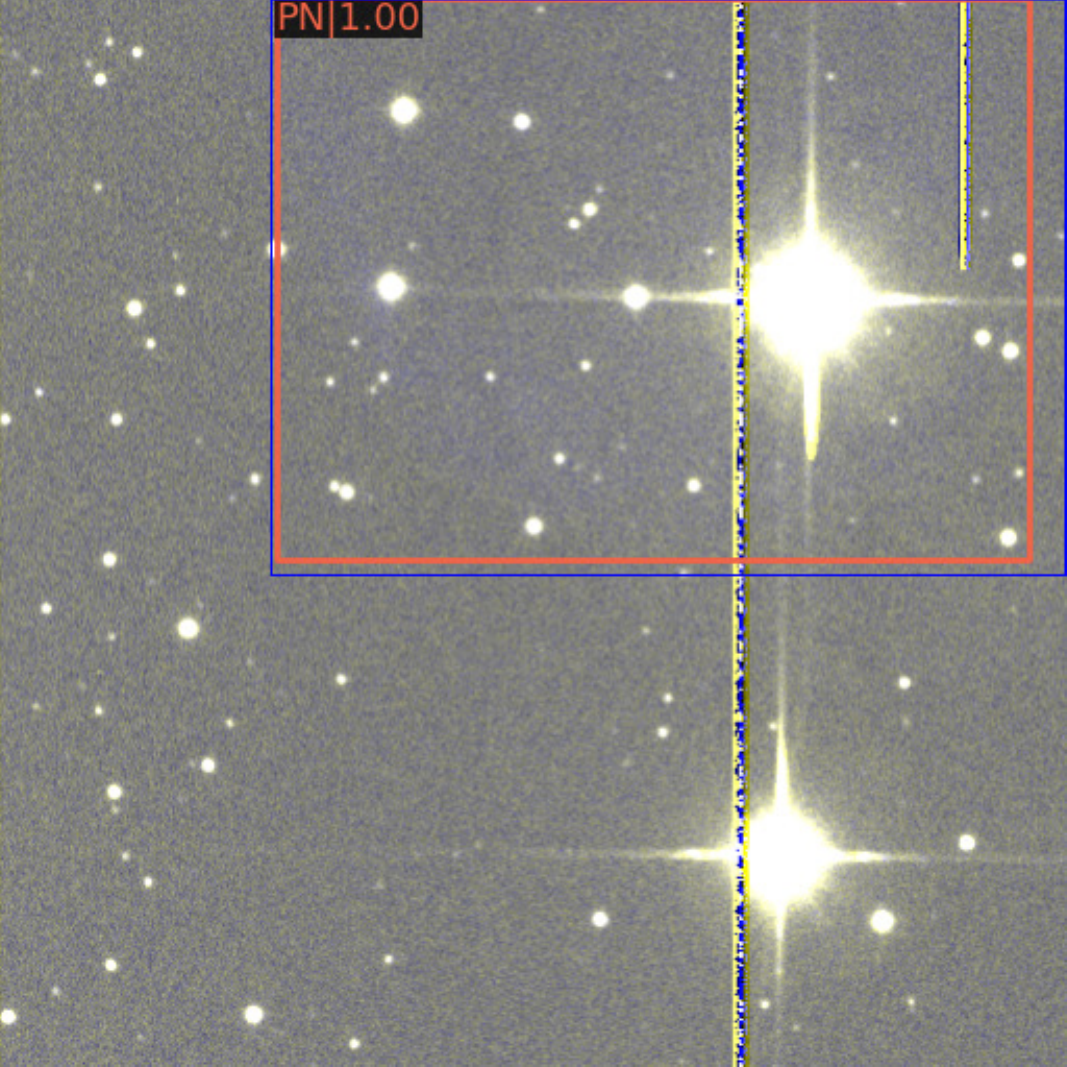}
\includegraphics[width=0.5\columnwidth]{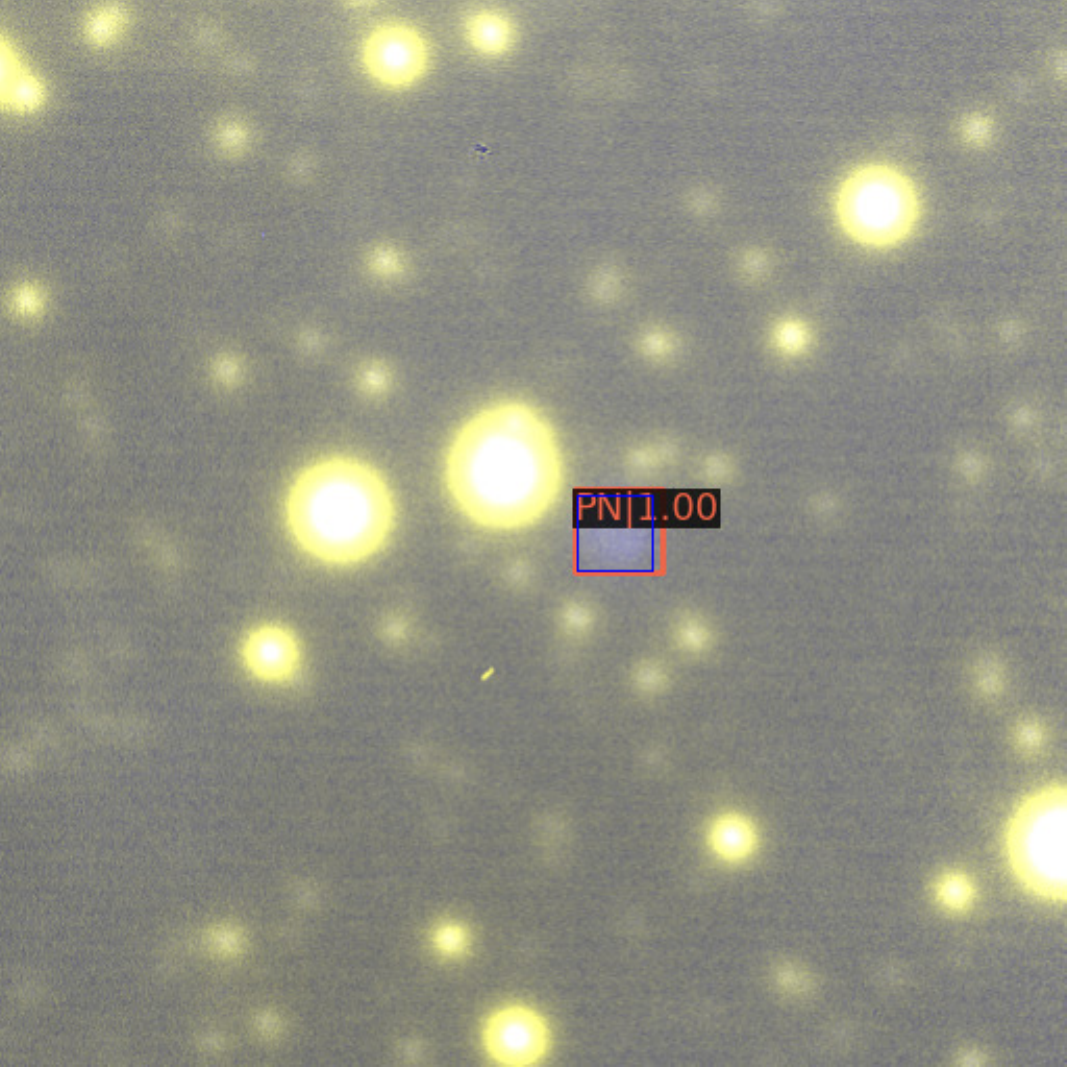}
\caption{Part of the detection results of the IPHAS validation set. As shown in this figure, our detection 
algorithm accurately identifies the targets and correctly determines their positions.}
\label{fig:results}
\end{figure}

We visualized the detection results of the model, as shown in the Figure \ref{fig:results}. The blue color represents the 
known positions of the PNe, while the red color represents the detected positions of the PNe, and the numbers in the 
labels indicate the confidence of the detections. The model not only performs extremely well in detecting PNe of different 
scales and morphologies but is also able to locate them in complex environments like dense star fields, contrasting 
extended emission regions or near bright stars. Furthermore, our techniques are effectively immune to effects of the 
instrument and observation conditions such as CCD issues and variations in seeing.

\begin{figure*}
\hspace*{-0.3in}
\includegraphics[width=0.9\textwidth]{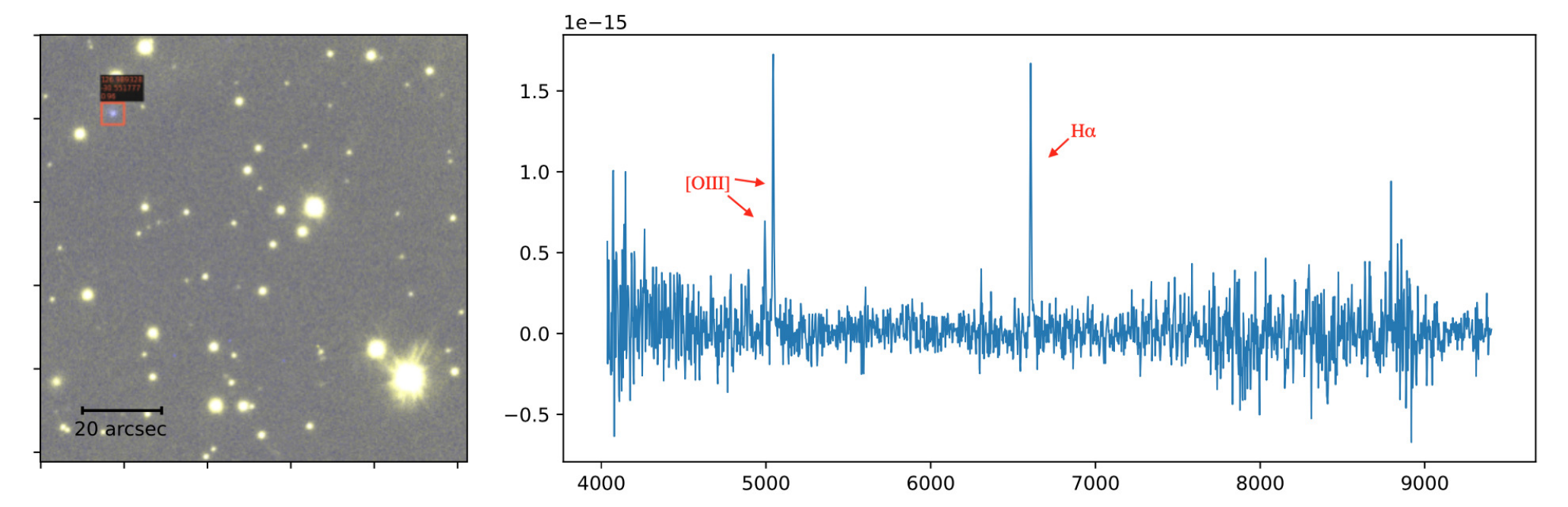}
\hspace*{-0.25in}
\includegraphics[width=0.9\textwidth]{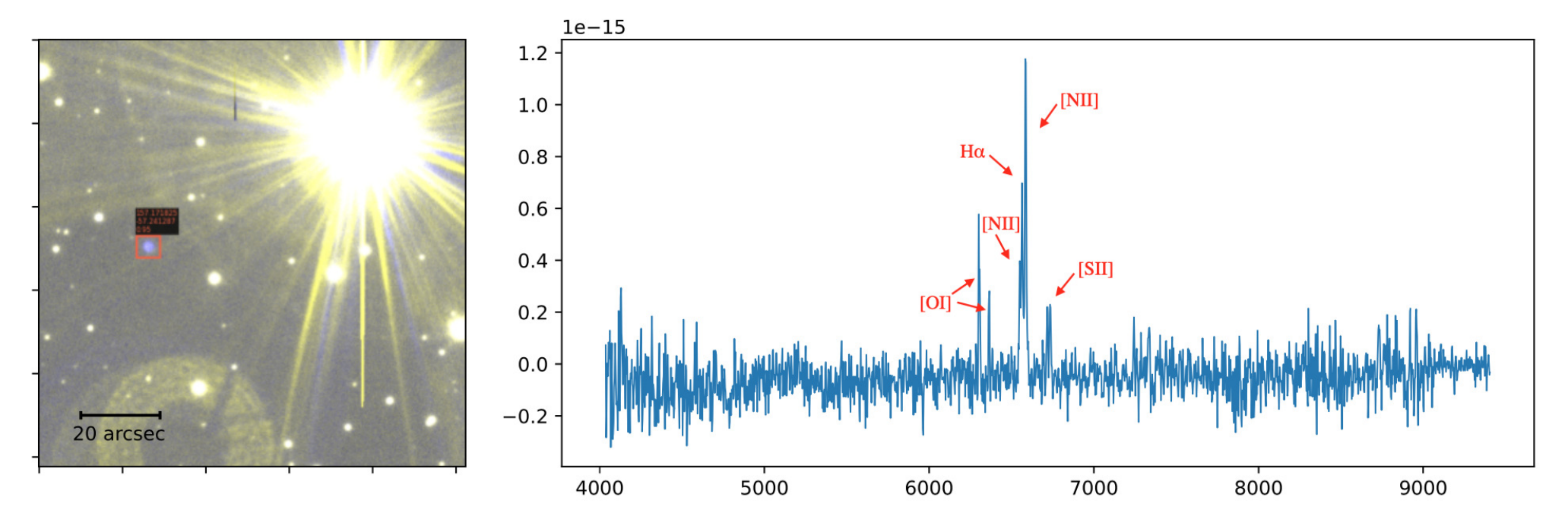}
\hspace*{-0.27in}
\includegraphics[width=0.9\textwidth]{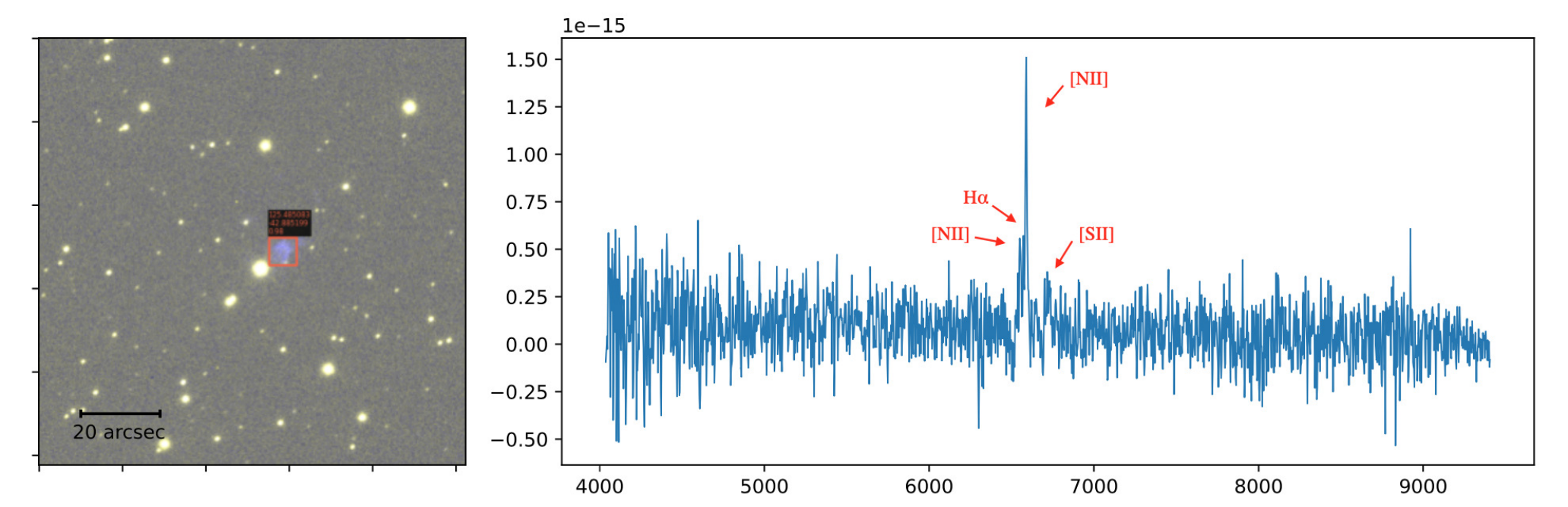}
\hspace*{-0.27in}
\includegraphics[width=0.9\textwidth]{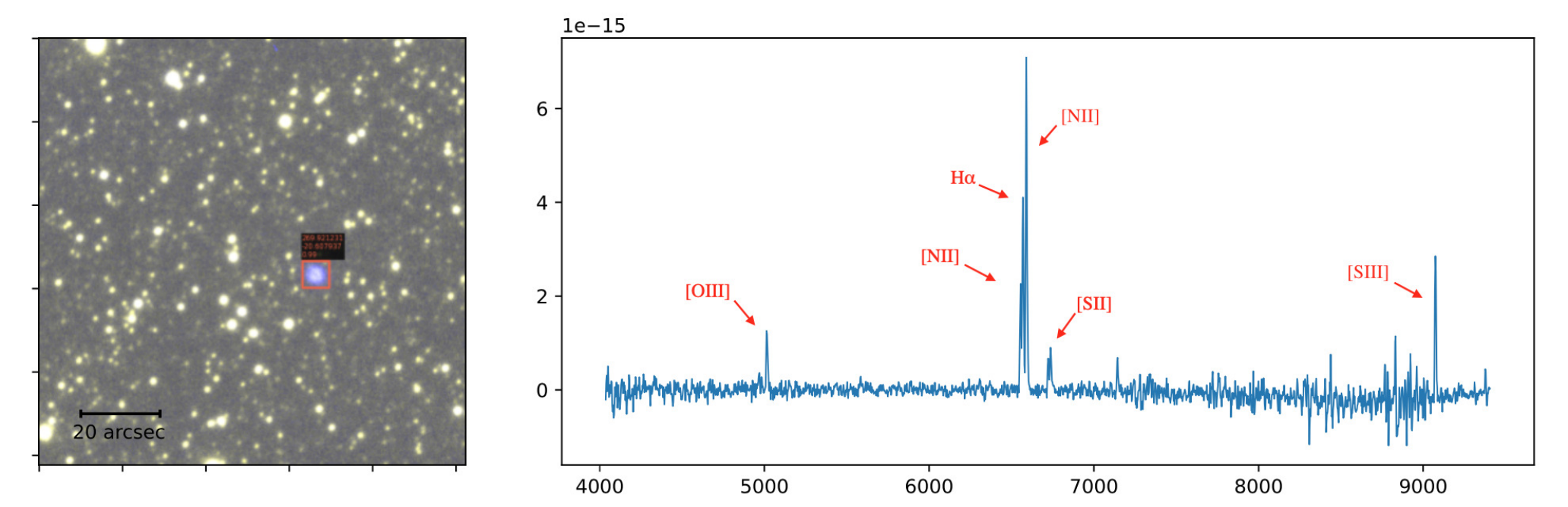}
\caption{Example PNe candidate discovery images from application of our ML process to the 
VPHAS+ survey data together with their 1-D SAAO 1.9m SpupNic confirmatory spectra classified, from top to bottom, 
as True, Likely, True and True PNe respectively following standard HASH processes. The PNe candidate is located 
within the red box in each image and has a blueish tinge due to the way the H$\alpha$ and red bands are combined. 
The common scale of the VPHAS+ images is 105 arcsec on 
each side. These form part of the larger sample spectroscopically confirmed and detailed 
in Paper~II. The bottom 3 of the 4 spectra have the [NII] emission lines stronger than H$\alpha$ at 6563\AA~ (a 
good diagnostic that eliminates HII region contaminants) but also weak or absent [OIII] in the blue due to 
extinction. The newly discovered targets tend to be relatively small and are frequently obscured or partially 
obscured by bright sources or in dense stellar regions. }
\label{fig:results of v}
\end{figure*}

\subsection{False negatives and positives from the IPHAS training data}
As mentioned from 454 known PNe in the IPHAS validation set we successfully
identified 444. Hence 10 (2.2\%) were not recovered by our ML technique. On examination it was found that they are largely due to the variable quality of some of the IPHAS imagery so the object is fainter and harder to find. In some cases larger diameter PNe were broken into smaller segments leading to a false negative for the object itself but several false positives.  We had 16 such false positives. These can also arise from diffuse scattered light effects around bright stars that can mimic emission as the broad and narrow band filters do not produce the same scattered light distribution. Some of these also arise because the algorithm mistakenly identifies other types of diffuse radiation as PNe, resulting in false positives. Some late type stars have very strong molecular bands that fall in the H-alpha band pass but are much weaker in the broad band red filter to the blue side of the molecular band. Both filters cut-off further to the red so the even stronger molecular bands in this region from these types of star are not sampled. This can give the effect of an apparent very strong H-alpha excess compared to the red broad band data and so a false positive.
\subsection{Results from VPHAS+}
\label{sec:results in the VPHAS+}

We used the trained model to scan 90,000 VPHAS+ observational data samples covering an area of 979 square 
degrees (43\% of the total survey footprint), which consists of approximately 4.5 million PNG images processed 
as described in Section \ref{sec:dataset}. Scanning one image of size $512\times 512$ takes 0.05 seconds. 
We visualized the output results, as shown in the Figure \ref{fig:results of v}. It can be seen that the model 
can achieve accurate detection for the majority of nebulae. Even emission regions which are obstructed or affected 
by nearby stars could be effectively revealed by our method. The model returned $\sim$20,000 detections. 
Then we compared the coordinates of these detections with the existing objects catalogue from HASH database with 
the criteria of the differences in image centroids in RA and Dec both smaller than 20 arcsec, separately for true PNe, 
likely \& possible PNe, and other kinds of known nebulae and emission-line objects. After that, we inspected the 
images visually, for diffraction spikes from bright stars, CCD issues  (such as bad pixels, defects etc), confusion by normal 
stars, very diffuse emissions, moving objects (between the epochs of the different exposures), very faint 
detections, as well as PNe candidates. After the above filtering and screening, we 
eventually identified 3452 candidate objects (17.3\%), including 2637 known 
nebula targets and 815 newly discovered high-quality candidates. Note that because the detections of algorithm 
can be verified by the known catalogue at a level that is difficult for the human eye to discern, it is highly 
possible that some faint targets are hiding in the "very faint detections" category.

\begin{figure}
\includegraphics[width=0.45\columnwidth]{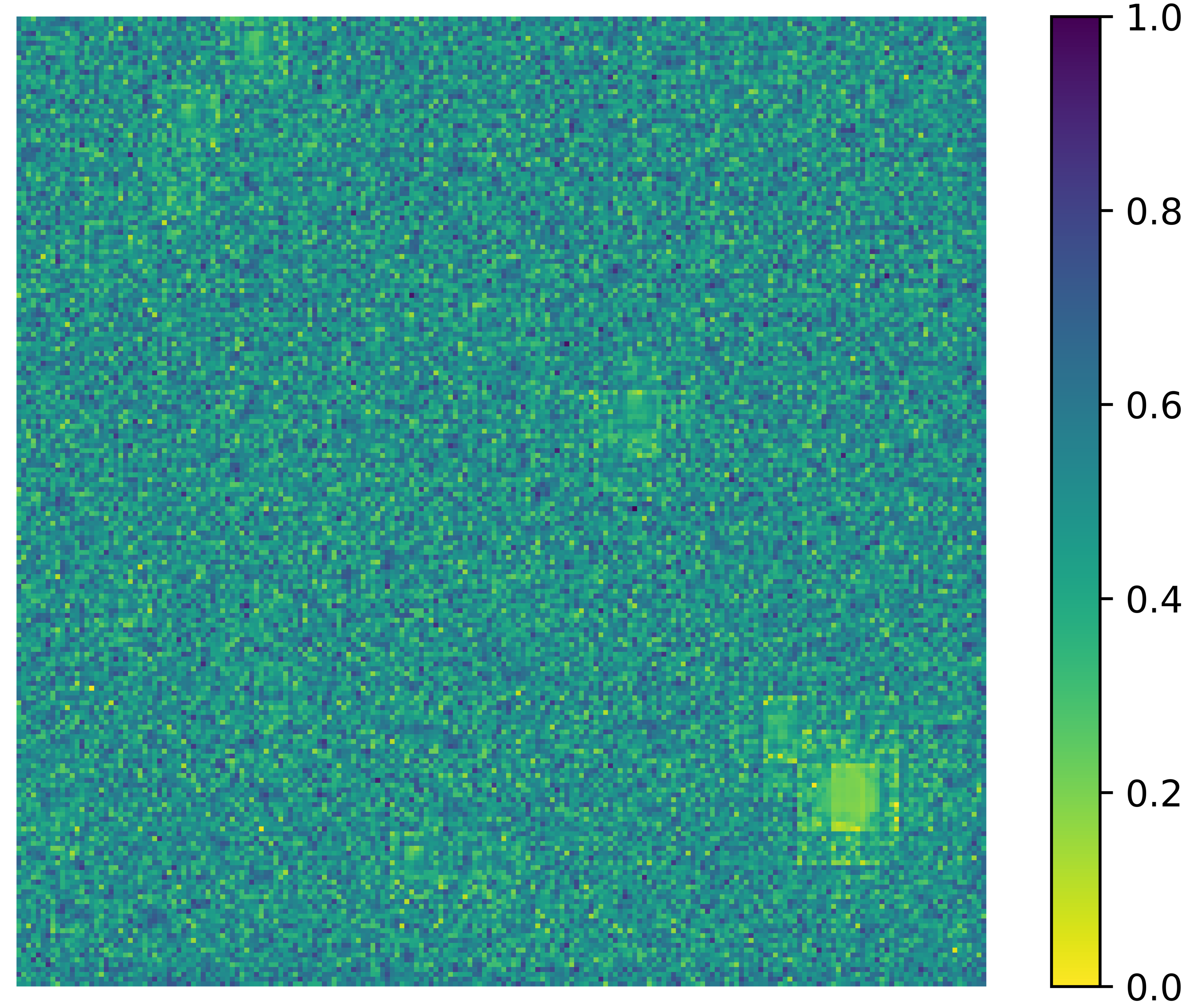}
\includegraphics[width=0.45\columnwidth]{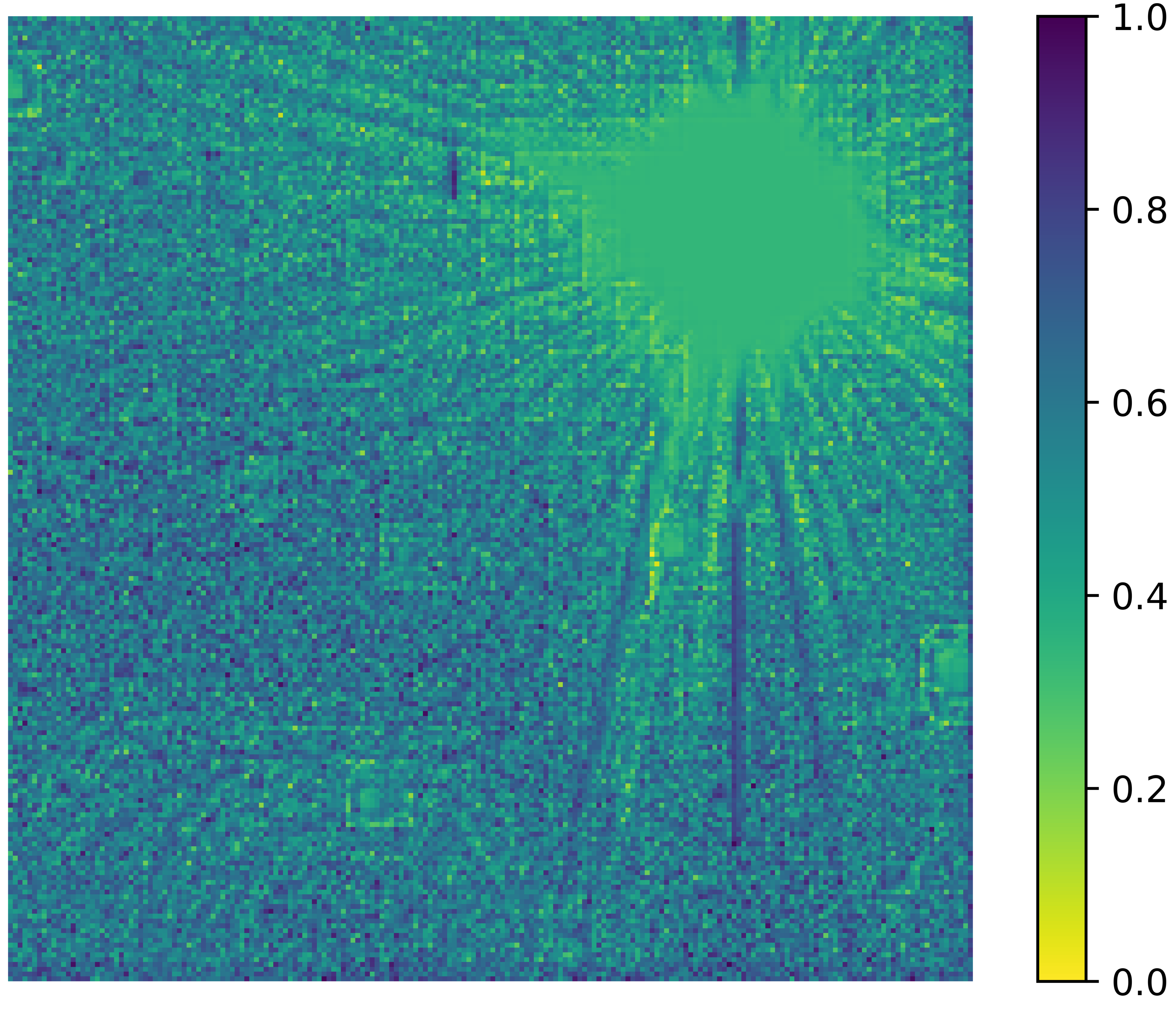}

\includegraphics[width=0.45\columnwidth]{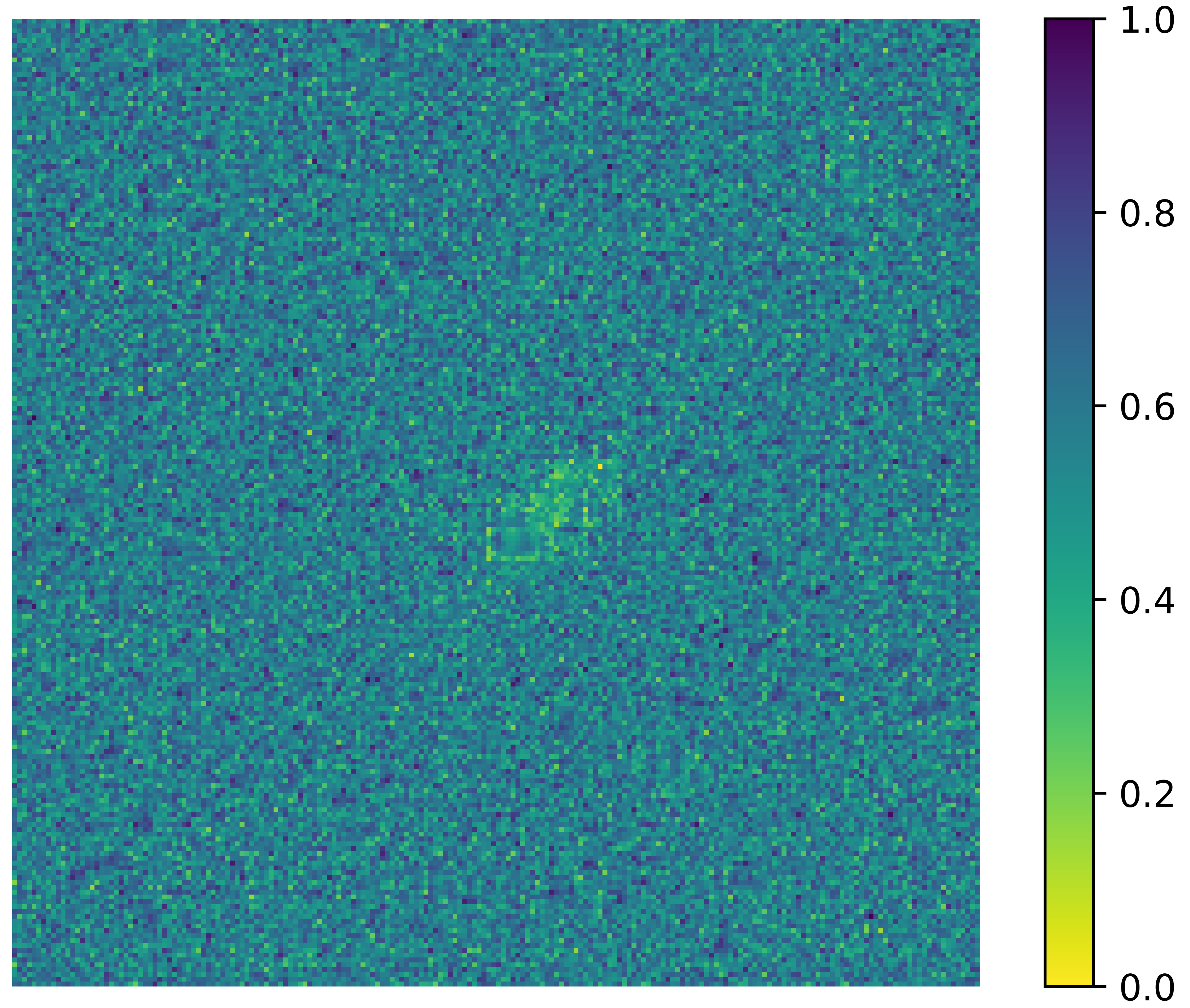}
\includegraphics[width=0.45\columnwidth]{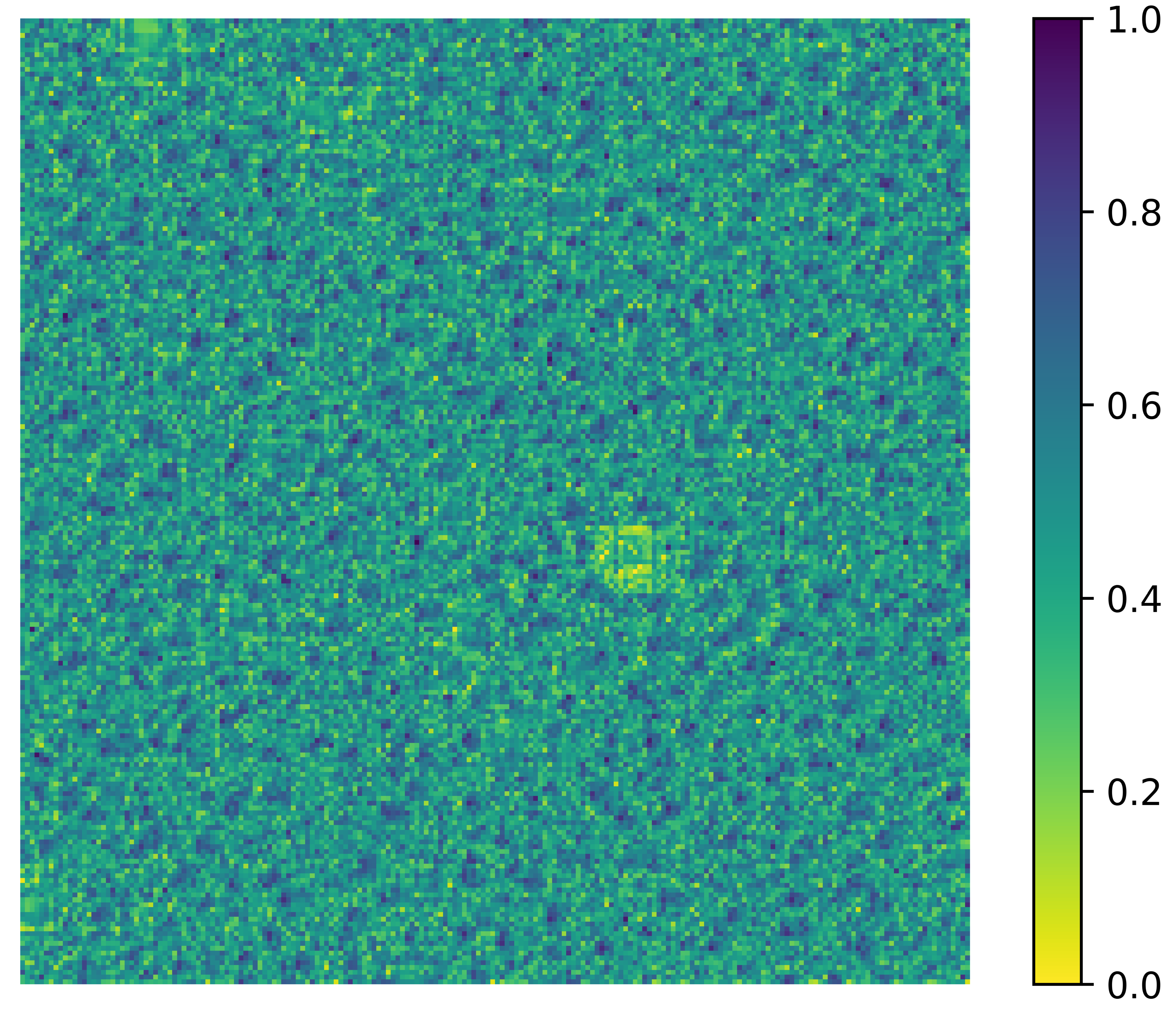}
\caption{The distribution of attention in the model for the four newly discovered targets corresponding to those in Figure 
\ref{fig:results of v}. It can be seen that in the feature extraction stage, the model has achieved preliminary 
object detection, filtered out most irrelevant targets, and retained nebula targets and a few interference targets 
with similar features, which can be removed by subsequent classification and regression modules.}
\label{fig:attention map}
\end{figure}

\section{Sample Analysis and preliminary spectroscopic results}
\label{sec:sample analysis}
We conducted further screening of the model's newly discovered high-quality PNe candidates through visual 
inspection by their morphology and double checked with SHS images.
We selected 31 of the most obvious candidates for follow-up spectroscopic confirmation on the 
SAAO 1.9~m telescope in July 2023. 

Detailed results and analysis
from this first spectroscopic observing program will be given in Yushan Li et al., Paper~II but we present 
here the VPHAS+ images and reduced 1-D spectra of four of the newly discovered, spectroscopicaly confirmed PNe 
in Figure \ref{fig:results of v}. Also in Table \ref{tab:num} we present the summary statistics from our 
preliminary spectroscopy. We find that 74\% of our candidates are confirmed PNe (either True (16), Likely (3) or 
Possible (4) following standard HASH prescriptions) and that 90\% have emission lines.  Only 3 out of 31 
$\sim10\%$ candidates are late-type star contaminants. These are extremely encouraging results for the scientific 
potential of this work.

\begin{table}
\centering
\caption{Summary statistics from our preliminary spectroscopy.}
\label{tab:survey}
\begin{tabular}{ |p{3cm}||p{1cm}|p{1cm}|p{1cm}|  }
 \hline
 \multicolumn{4}{|c|}{Total SAAO 1.9~m spectroscopically observed targets: 31} \\
 \hline
 Classification & True & Likely & Possible \\
 \hline
 Planetary Nebula & 16 & 3 & 4 \\
 Supernova Remnant & 0 & 0 & 1 \\
 Emission-line Star & 2 & 0 & 0 \\
 Late-type Star & 4 & 0 & 0 \\
 Emission in star cluster & 1 & 0 & 0 \\
 \hline
 \label{tab:num}
\end{tabular}
\end{table}

The discovery images and 1-D spectra of 4 detected PN candidates are shown in 
Figure \ref{fig:results of v} as an illustrative example of the success of our technique. 
The 107$\times$107~arcsecond VPHAS+ image of the PN candidate is shown on the right. The PNe
candidate is located within the red box in each image and has a blueish tinge due to the way the H$\alpha$ and 
red bands are combined. The SAAO 1.9~m 1-D spectra for this PNe candidate is shown on the left covering the 
wavelength range from $\sim$4000$\AA$ to 9500$\AA$. The standard PN emission lines of
H$\alpha$, [N II] and [S II] are very clear in the red in the middle of the spectral plots. The bottom 3 of the 
4 spectra have the [NII] emission lines stronger than H$\alpha$ at 6563\AA (a good
diagnostic that eliminates HII region contaminants) but also weak or absent [OIII] in the blue due to extinction. 
These spectra confirm the PN nature of the candidates and verify the power of our technique to find new PNe.


Furthermore, we visualize the attention distributions of these targets in the model, revealing 
the important features extracted by the model as shown in Figure \ref{fig:attention map}. 
It can be seen that around some special celestial bodies, including PNe, there is a significant 
difference in attention distribution from the surrounding background. Thanks to the 
attention mechanism, our model has already achieved preliminary object filtering in the feature 
extraction stage, and this filtering is at the pixel level, which far exceeds the accuracy 
requirements of object detection and is a rough object segmentation. This helps deepen our 
understanding of PNe as a diverse object class in the data. 

\section{Conclusions and Prospects}
\label{sec:conclusions and prospects}
In this paper, we use the Swin-transformer object detection model in searching for
PNe in H$\alpha$ surveys of the inner Galactic plane. Compared to traditional machine learning 
convolutional neural network object detection models, this object detection model, based on an 
attention mechanism, can ignore the restriction of receptive field to model any range of features, 
so it has better universality and detection effect capability for objects of different spatial scales. 
Meanwhile, compared to the ordinary transformer model, Swin-transformer model can effectively improve 
the defect of high model complexity and large calculation, so as to have better adaptability.

We trained and validated the model using the known PNe catalogue and images from IPHAS. 
After training, the model reached an accuracy rate of $96.5\%$ and a recall rate of $97.8\%$ 
on the verification set, which has excellent detection ability. We further applied the trained model 
to the VPHAS+ processed images and obtained $\sim$20,000 detections, which, by comparing with existing 
catalogues, resulted in 2,637 known PNe and 815 new high-quality candidates.

Through visual inspection, we selected the most promising 31 candidates for confirmatory spectroscopic observations. 
A summary table of the classification results from these observations is presented in Table.2
We also provide  four examples of new PNe uncovered by our ML techniques together with 
their confirmatory spectroscopy. These preliminary findings demonstrates the strong performance of our model 
as a powerful tool for efficiently discovering nebula targets in wide-field H$\alpha$ survey imagery.  Full details are 
presented in Yushan Li et al. (paper~II, in preparation). A more comprehensive and detailed search for nebula targets and 
PNe candidates in VPHAS+ data is also underway and will be presented in future work.

Finally, our attention-based object detection model not only performs well 
in detecting PNe candidates, discovering many new objects, but has the potential to be adapted to 
other astronomical targets anfd to other large-scale sky surveys in the future.

\section*{Acknowledgements}

This work is supported by National Natural Science Foundation of China (NSFC) with funding number of 12303105, 12173027 
and 12173062 and Civil Aerospace Technology Research Project (D050105). We acknowledge the science research grants from 
the China Manned Space Project with NO. CMS-CSST-2021-A01 and science research grants from the Square Kilometre Array 
(SKA) Project with NO. 2020SKA0110102. QAP thanks the Hong Kong Research Grants Council for GRF research support under 
grants 17326116 and 17300417. YL thanks HKU and QAP for provision of a PhD scholarship from RMGS funds awarded to the LSR.

\section*{Data Availability}

Data resources are supported by China National Astronomical Data
Center (NADC) and Chinese Virtual Observatory (China-VO). This
work is supported by Astronomical Big Data Joint Research Cen-
ter, co-founded by National Astronomical Observatories, Chinese
Academy of Sciences and Alibaba Cloud. Once this paper is accepted, the code and data utilized 
can be accessed from the PaperData Repository, managed by the China-VO team.



\bibliographystyle{mnras}
\bibliography{example} 








\bsp	
\label{lastpage}
\end{document}